\documentclass[twocolumn,showpacs,pra,aps,superscriptaddress,amssymb]{revtex4-1}
\usepackage{amsfonts}
\usepackage{mathrsfs}
\usepackage{amssymb}
\usepackage{bm}
\usepackage{graphicx}
\usepackage[colorlinks=true,linkcolor=blue,citecolor=blue,urlcolor=blue]{hyperref}

\begin{document}

\title{Nonexponential tunneling decay of a single ultracold atom}
\author{Gast\'on Garc\'{\i}a-Calder\'on}
\email{gaston@fisica.unam.mx}
\affiliation{Instituto de F\'{\i}sica, Universidad Nacional Aut\'onoma de M\'exico,
Apartado Postal {20 364}, 01000 M\'exico, Distrito Federal, Mexico}
\author{Roberto Romo}
\email{romo@uabc.mx} \affiliation{Facultad de Ciencias, Universidad
Aut\'onoma de Baja California, Apartado Postal 1880, 22800 Ensenada,
Baja California, Mexico}
\date{\today}

\date{\today}

\begin{abstract}
By using an exact analytical approach to the time evolution of decay we investigate the tunneling decay of ultracold single atoms, to discuss the conditions for deviations of the exponential decay law. We find that  $R$, given by the ratio of the energy of the decaying fragment $\mathcal{E}_r$ to its corresponding width $\Gamma_r$, is the relevant quantity in this study. When $R$ is less than $0.3$ the decay of the atom goes to a good approximation for the first few lifetimes as $\exp(-\Gamma_rt/2\hbar)t^{-3/2}$. We also find that for values of $R \sim 1$, the  nonexponential behavior occurs in a post-exponential regime that goes as $t^{-3}$ after around a dozen of lifetimes.  The  above conditions depend on suitable designed potential parameters and suggest that for values  $R \lesssim 1$, the experimental verification of nonexponential decay might be possible.
\end{abstract}


\maketitle
\section{Introduction}
The recent experimental work on tunable few-fermion systems consisting of ultracold gases in optical traps, that is characterized by the control of the quantum state of the system \cite{jochim11,zurn12,zurn13a}, has opened the way to investigate a variety of aspects of few and many-body physics \cite{bloch08,lode12,guan13}. As described in these works, this can be achieved by exploiting Pauli's principle in a highly degenerate Fermi gas in a trap so that it is possible to control the number of particles by controlling the number of available lowest energy single-particle states. For a few particles, the confining trapping potential consists of a one-dimensional optical potential created by a tight focus of a laser beam and a magnetic field gradient in the axial direction, in such a way that the states above a well defined energy become unbound. The resulting potential is formed by an impenetrable barrier on the left, a barrier of finite height on the right and a well in between with a controlled number of atoms which may decay out of the trap by tunneling through the barrier. The above setup has been used to address experimentally the tunneling decay of two or more atoms \cite{jochim11}, and has motivated theoretical studies on the dynamics of multiparticle decay \cite{cdgcmr06,delcampo11,lode12,longhi12,pons12}, in addition to studies on the decay of two interacting or non-interacting particles \cite{gcm11,kim11,zurn12,rontani12,rontani13,lundmark15,gharashi15}. In Ref. \cite{jochim11} it is also pointed out the possibility to prepare just one atom in the lowest energy level of an optical trap.

In contrast to the widespread view that tunneling decay of an isolated single particle into open space is amply understood
\cite{lode12}, here we call the attention to the old prediction of the deviations of the exponential decay law at short and long times compared with the lifetime of the decaying system. Here, by using an exact analytical approach for decay \cite{gc10,gc11,gcmv12}, we investigate the conditions that need to be fulfilled to be able to observe nonexponential decay of single ultracold atoms tunneling out of  potential profile using realistic parameters. We believe that these systems are the closest realization for tunneling decay in a highly isolated environment and hence might be appropriate to test nonexponential decay at long times. This is a relevant issue from a fundamental point of view that requires experimental verification.

We first provide an overview of the subject of decay of particles by tunneling. It is well known that quantum decay, a subject as old as quantum mechanics, was developed to explain $\alpha$-decay in radioactive nuclei. In 1928, Gamow derived the analytical expression for the exponential decay law $\exp(-\Gamma t/\hbar)$, with $\Gamma$ the decay rate; an expression that has been widely used in the description of particle decay. For single-particle decay, and it seems also in multi-particle decay \cite{delcampo11,pons12}, one may usually identify, in addition to the exponential regime,  two nonexponential regimes that, respectively, occur in general at short and long times compared with the lifetime of the system. The short-time behavior, which is related to the existence of the energy moments of the Hamiltonian \cite{khalfin68}, exhibits typically a $t^2$ behavior (see however \cite{cgc12}) and has been the subject of a great deal of attention, particularly in connection with the quantum Zeno effect \cite{sudarshan77b,koshino05}. The long-time, post-exponential, regime is a consequence of the fact that in most real systems, the energy spectra $E$ is bounded by below, i.e., $E \in (0,\infty)$, leading to integer inverse power in time behaviors, as discussed by Khalfin \cite{khalfin58}. In that work, Khalfin indicated the relevant role of the ratio of the energy of the decaying fragment $\mathcal{E}_r$ to the decaying width $\Gamma_r$, $R=\mathcal{E}_r/\Gamma_r$, in determining the time scale for the transition from  exponential to  nonexponential decay. Subsequent theoretical work investigated further the issue of the approximate nature of the exponential decay law \cite{goldberger64,baz69} and  provided also estimates of the above time scale for values  $R \gg 1$ \cite{winter61,ghirardi78,peres80}. This is of interest because it gave an explanation of the failure of finding deviations of the exponential decay law  at long times in radioactive nuclei \cite{norman88,son98}. Norman et.al. \cite{norman88} looked unsuccessfully for deviations from exponential decay law using $^{56}{\rm Mn}$ up to $65$ lifetimes. Here $R \sim 10^{17}$ and Winter's estimate in lifetime units: $t_0=5\ln(R)$ for the onset of nonexponential decay, yields $t_0 \sim 200$, which is beyond the experimental range of present day technology. Following the work by Khalfin \cite{khalfin58}, it was soon realized that deviations from the exponential decay law could also been obtained for small values of $R$, as the proposal for observing nonexponential decay in isolated autoionizing states located very close to the energy threshold in atomic systems \cite{nicolaides77}, that so far has not been confirmed experimentally.

The above results contributed to the widespread view that nonexponential decay contributions were beyond experimental reach or even to the alternative explanation that the interaction of the decaying system with the environment would enforce exponential decay at all times \cite{ghirardi78,parrot02}. Some years ago, however, short-time deviations from exponential decay \cite{raizen97} and the quantum Zeno effect \cite{raizen01} were finally observed, and more recently, in 2006, the measurement of post-exponential decay in a number of organic molecules in solution, exhibiting distinct  inverse power in time behaviors was reported \cite{monk06}. The above long-time deviations from the exponential decay law, however, were obtained due to the additional broadening of the excited state energy distributions produced by the solvent. That is, instead of looking for states with very small $E$, these authors considered systems with large values of $\Gamma$. It is intriguing that the solvent, which may be regarded as some sort of environment for these complex molecules, favors a nonexponential behavior. To the best of our knowledge, there is not at present time a theoretical approach that explains the above experimental results. It is worthwhile to point out that these experiments refer to luminescence decays after laser pulse excitation and hence do not refer to particle decay as considered in the present work.

In 2005 Jittoh et. al. published a theoretical study on particle decay using a single-pole approximation, where these authors estimate that for values of $R < 0.5$  in $s$-wave spherical symmetric systems or one-dimensional systems, there exists a novel regime where the decay is nonexponential at all times \cite{jittoh05}. These authors, however, did not discuss that regime in actual physical systems. In 2006, Garc\'{\i}a-Calder\'on and Villavicencio  \cite{gcv06} suggested the possibility that this novel full time nonexponential regime could be observed in semiconductor double-barrier resonant quantum structures.

As pointed out at the beginning of this Introduction, in this work we investigate the conditions that need to be fulfilled to be able to observe nonexponential decay in the deterministic preparation of a tunable single atom in an optical trap \cite{jochim11}. As pointed out above, we believe that these systems are the closest realization  of  decay by tunneling of a particle out of a single particle potential. We derive  analytical expressions for the nonescape probability as an expansion involving the full set of decaying states of the system at all times and study the conditions of validity for the single-pole approximation.

It is worth to point out that the formulation considered here refers to the full hamiltonian $H$ to the problem and hence it differs from approaches where the Hamiltonian is separated into a part $H_0$ corresponding to a closed system and a part $H_1$ responsible for the decay which is usually treated to some sort of perturbation theory, as in the work by Weisskopf and Wigner to describe the decay (also exponential) of an excited atom interacting with a quantized radiation field \cite{wigner30a}. These approximate approaches have become a standard procedure for treating a class of decay problems where perturbation theory can be justified, as in studies of  nonexponential decay in atomic spontaneous emission \cite{mostowski73,knight77}.

The manuscript is organized as follows. Section II provides an overview of the theoretical formalism that we consider here. Section III discusses some model calculations and analyzes different post-exponential scenarios and finally, Section IV presents some Concluding Remarks.

\section{Formalism}

The formalism that we shall consider here has its roots in the old work by Gamow which imposed outgoing boundary conditions on the solutions to the Schr\"odinger equation to describe the process of decay \cite{gamow28,gamow49}. As is well known, these boundary conditions lead to complex energy eigenvalues, its imaginary part being twice the decay rate that appears in the expression of the exponential decay law. Outside the interaction region, the amplitude of such solutions, known as decaying, resonant or quasi-normal states, grows exponentially with distance  and hence the usual rules of normalization and completeness do not apply. The approach initiated by Gamow, however, evolved over the years. In particular, significant developments in the 1970s on the analytical properties of the outgoing Green's function to the problem provided a suitable framework to study distinct  approaches to the issues of normalization and eigenfunction expansions involving these states \cite{berggren68,gcp76}. In particular some of these developments have led to an exact analytical description of decay by tunneling \cite{gc10,gc11}.

The effective trap potential  that results after application of the magnetic field to the initially confining trap and of the spilling process that guarantees that the decaying atom remains in the lowest decaying state,
corresponds to a one-dimensional system where the  transmission channel is closed \cite{jochim11,rontani13}. This potential is analogous to a spherical potential of zero angular momentum.

The solution to the time-dependent Schr\"odinger equation as an initial
value problem, may be written at time $t$ in terms of the retarded Green's function $g(x,x';t)$ of the problem as \cite{gc10,gc11}
\begin{equation}
\Psi(x,t)=\int_0^L {\! g(x,x^\prime,t)\Psi(x',0)\,\mathrm{d}x^\prime},
\label{e1}
\end{equation}
where $\Psi(x,0)$ stands for a state initially confined within the internal interaction region
$(0,L)$. Here, for simplicity of the discussion and without loss of generality it is assumed that $\psi(x,0)$ is a real function.
A convenient form of the retarded time-dependent Green's function is expressed in terms of the
outgoing Green's function $G^+(x,x';k)$ of the problem. Both quantities are related by a Laplace transformation
\cite{gc10,gc11}. In the present approach, instead of the common practice of assuming the analytical properties of $G^+(x,x';k)$, we impose the condition, justified on physical grounds, that the potential vanishes after a distance, i.e. $V(x)=0, \,x>L$.  As a consequence, it can be rigorously proved that $G^+(x,x';k)$  may be extended analytically to the whole complex $k$ plane where it has an infinite number of poles  distributed in a well known manner \cite{newtonchap12}.

The relevant point here is that the residue of $G^+(x,x';k)$ at a pole $\kappa_n$
is proportional to the functions $u_n(x)$ and $u_n(x^{\prime})$ and provides its normalization condition \cite{gcp76,gc10,gc11}.
The  decaying or resonant states $u_n(x)$ satisfy the Schr\"{o}dinger equation of the problem
\begin{equation}
[E_n-H]u_n(x)=0,
\label{en1}
\end{equation}
where $H$ is the full Hamiltonian $H=-(\hbar^2/2m)d^2/dx^2 + V(x)$, with $m$ the mass of the decaying particle. Equation (\ref{e1}) satisfies outgoing  boundary conditions at $x=L$, namely,
\begin{equation}
u_n(0)=0,\qquad \left[\frac{du_n(x)}{dx}\right]_{x=L}=i\kappa_n u_n(L),
\label{en2}
\end{equation}
with $\kappa_n=\alpha_n-i\beta_n$. The quantity $E_n$ in (\ref{en1}) refers to the complex energy eigenvalue  $ E_n= (\hbar^2/2m)\kappa_n^2=\,\mathcal{E}_n-i\Gamma_n/2$, where $\mathcal{E}_n$ yields the resonance energy of the decaying fragment and $\Gamma_n$ stands for the corresponding decaying width. Using Cauchy's Integral Theorem allows  to obtain a discrete expansion of $G^+(x,x^{\prime};k)$ in terms of the functions $\{u_n(x)\}$ and the poles $\{\kappa_n\}$ along the internal potential region. This expansion may be used to obtain a representation of $g(x,x^{\prime},t)$ that may be inserted into Eq. (\ref{e1}) to obtain the time-dependent solution \cite{gc10,gc11}
\begin{equation}
\Psi(x,t)=\sum_{n=-\infty}^{\infty}
\left \{ \begin{array}{cc}
C_nu_n(x)M(y^\circ_n), & \quad  x \leq L \\[.3cm]
C_nu_n(L)M(y_n), & \quad  x \geq  L,
\end{array}
\label{e2}
\right.
\end{equation}
where the coefficients $C_n$ are given by
\begin{equation}
C_n=\int_0^L \Psi(x,0) u_n(x) dx,
\label{3c}
\end{equation}
and the functions $M(y_n)$ are defined as \cite{gc10}
\begin{eqnarray}
M(y_n)&=&\frac{i}{2\pi}\int_{-\infty}^{\infty}\frac{{\rm e}^{ik(x-L)}{\rm e}^{-i\hbar k^2t/2m}}{k-\kappa_n}dk\nonumber \\ [.3cm]
&&=\frac{1}{2}{\rm e}^{(imr^2/2 \hbar t)} w(iy_n),
\label{e3}
\end{eqnarray}
where $y_n={\rm e}^{-i\pi /4}(m/2\hbar t)^{1/2}[(x-L)-(\hbar \kappa_n/m)t]$, and the function $w(z)=\exp(-z^2)\rm{erfc(-iz)}$ stands for the Faddeyeva or complex error function \cite{abramowitzchap7} for which there exist efficient computational tools \cite{poppe90}. The argument $y_n^{\circ}$ of the functions $M(y_n^0)$ in (\ref{e2}) is that of $y_n$ given above with $x=L$.

Notice that the sums in (\ref{e2}) run, respectively, over the poles  $\kappa_{-n}= -\alpha_n -i \beta_n$, located on the third quadrant of the $k$ plane, and the poles $\kappa_{n}= \alpha_n -i \beta_n$, located on the fourth quadrant. It follows from time reversal invariance that $\kappa_{-n}=-\kappa_n^*$ \cite{newtonchap12}.

The functions $\{u_n(x)\}$ are normalized according to the condition
\begin{equation}
\int_0^L u_n^2(x) dx + i\frac{u_n^2(L)}{2\kappa_n}=1,
\label{norm}
\end{equation}
and satisfy a closure relationship along the internal region of the potential
which,  provided the initial state is normalized to unity, leads to the expression \cite{gc10,gc11},
\begin{equation}
{\rm Re}\left \{ \sum_{n=1}^\infty C^2_n \right\}= 1,
 \label{e5}
\end{equation}
Equation (\ref{e5}) indicates that  the terms ${\rm Re}\,\{C^2_n\}$ cannot be interpreted as a probability, since in general they are not positive definite quantities, however, each of them  represents the `strength'  or `weight' of the initial state in the corresponding decaying state. One might see the coefficients  ${\rm Re}\,\{C^2_n\}$ as some sort of quasi-probabilities \cite{halliwell13}.

The equivalence between the non-Hermitian formulation that leads to the time-dependent solution given by Eq. (\ref{e2}) and the  Hermitian formulation based on continuum wave functions is discussed in Ref. \cite{gcmv12}. There, the advantage of using the  analytical expressions for the distinct decaying regimes that follow from the former formulation is contrasted  with the `black box'  numerical treatment that characterizes the latter formulation.

It is worth mentioning that the formalism outlined above differs from the so called rigged Hilbert space formulation in many respects, as discussed in Refs. \cite{mgcm05,gc10}. For example, since in that approach the poles located on the third quadrant of the $k$ plane are not taken explicitly into consideration, there is no analytical description of the nonexponential contributions to decay, as given by the Eq. (\ref{e11}) discussed below. It might also be worthwhile to mention here that decaying states, in spite of its non-Hermitian nature, have been used in a large variety of topics yet with different names: resonant states, quasinormal modes or Siegert states, as for example in quantum transients \cite{gcr97,cgcm09}, gravitational waves and black holes \cite{starinets09} and nonadiabatic processes involving molecules \cite{tolstikhin08}.

\subsection*{Nonescape probability}

Two quantities of interest in decaying problems are the survival probability $S(t)$, that yields the probability that at time $t$ the system remains in the initial state, and the nonescape probability $P(t)$, that provides the probability that at time $t$ the particle remains within the confining region of the potential. When the initial state overlaps strongly with the lowest decaying state,  both quantities exhibit a very similar behavior with time \cite{gcmv07}.  It  seems that such is the case for atom decay in ultracold traps, where the spilling process leaves just one atom in the lowest decaying state \cite{jochim11}.

Here we consider the nonescape probability which is defined as
\begin{equation}
P(t) = \int_0^L \Psi^*(x,t)\Psi(x,t) \,dx.
\label{e8}
\end{equation}
In order to calculate the above quantity one  requires  the time-dependent solution along the internal region of the potential. Hence one may insert the top expression of Eq. (\ref{e2}) into Eq. (\ref{e8}),  to obtain the expansion of the nonescape probability in terms of decaying states,
\begin{equation}
P(t)=\sum_{m,n=-\infty}^{\infty}C_m C_n^*I_{mn}M(y_{m})M^*(y_{n}).
\label{pfull}
\end{equation}
where
\begin{equation}
I_{mn}= \int_{0}^{L} u_m(x) u^*_n(x) dx.
\label{imn}
\end{equation}
Equations (\ref{e2}) and (\ref{pfull}) are given in terms of $M$ functions, and consequently their exponential and nonexponential behavior is not exhibited explicitly. This may be obtained by using the symmetry relations $\kappa_{-n}=-\kappa_n^*$ and $u_{-n}(x)=u_n^*(x)$, to write the sums over the poles located on the fourth quadrant. Here one may use the relation $M(y^{\circ}_n)=\exp(-i\mathcal{E}_n t/\hbar)\exp(-\Gamma_n t/\hbar)-M(-y^{\circ}_n)$ \cite{gc10,abramowitzchap7}, to write the time-dependent wave function along the internal region as,
\begin{equation}
\Psi(x,t) = \sum_{n=1}^{\infty} C_nu_n(x){\rm e}^{-i\mathcal{E}_nt\hbar }{\rm e}^{-\Gamma_nt/2\hbar} - I_n(x,t), \quad x \leq L,
\label{e9}
\end{equation}
where the nonexponential contribution $I_n(x,t)$ is given by
\begin{equation}
I_n(x,t)= \sum_{n=1}^{\infty} C_nu_n(x)M(-y^{\circ}_n) - C^{*}_nu^{*}_n(x)M(y^{\circ}_{-n}).
\label{e10}
\end{equation}
In this last expression the argument $y^{\circ}_{-n}$ is equal to $y^{\circ}_{n}$ with $\kappa_{n}$ substituted by $\kappa_{-n}=-\kappa_n^*$.

Substitution of Eqs. (\ref{e9}) and (\ref{e10}) into Eq. (\ref{e8}) provides, therefore, an expression for the nonescape probability that exhibits explicitly the exponential and nonexponential contributions to decay.  Notice, that assuming an initial state that overlaps strongly with the longest lifetime state, say $n=r$, it may be seen  in view of (\ref{e5}), that ${\rm Re}\,\{C^2_r\} \approx 1$, and also that $I_{rr} \approx 1$, and ignoring  the nonexponential contributions, one obtains the well known exponential decay law $P(t)=\exp(-\Gamma t/\hbar)$.

Equations (\ref{e2}) and (\ref{pfull}) are exact and may be used to find out the validity of different approximations.

The functions $M(-y_{n}^{\circ})$ and $M(y_{-n}^{\circ})$ that appear in the nonexponential contribution given by Eq. (\ref{e10}), exhibit at long times a  $t^{-3/2}$  behavior with time \cite{gc10}. As a consequence, the time-dependent solution may be written along the exponential and long-time regimes as \cite{gc10},
\begin{eqnarray}
\Psi(x,t) &\approx & \sum_{n=1}^{\infty} C_nu_n(x){\rm e}^{-i\mathcal{E}_nt/\hbar}{\rm e}^{-\Gamma_nt/2\hbar} - \nonumber \\ [.3cm]
&&ib\,{\rm Im} \left\{\sum_{n=1}^{\infty}\frac{C_nu_n(x)}{\kappa_n^3} \right\} \,\frac{1}{t^{\,3/2}};\,\,x \leq L,
\label{e11}
\end{eqnarray}
with
\begin{equation}
b= \frac{e^{-i \pi /4}}{2 \sqrt{\pi}} \left(\frac{2m}{\hbar} \right)^{3/2}.
\label{b}
\end{equation}

\subsection*{Decay of a single level}

In what follows we restrict the discussion to the situation that corresponds to an atom located in the lowest decaying state, $n=1$,  of the effective ultracold trap potential. On physical grounds one expects that the initial state overlaps strongly with that state, and therefore  ${\rm Re} \{C^2_1\}$ may provide the main contribution to Eq. (\ref{e5}). This justifies to consider the single pole approximation, $n=1$, in the expansion of the decaying wave function given by Eq. (\ref{e11}). This is a good approximation except near the time origin where more poles are needed \cite{cgc12}. Hence we may write,
\begin{eqnarray}
\Psi(x,t) &\approx & C_1u_1(x){\rm e}^{-i\mathcal{E}_1t/\hbar}{\rm e}^{-\Gamma_1t/2\hbar} - \nonumber \\ [.3cm]
&&ib\,{\rm Im}\left \{\frac{C_1u_1(x)}{\kappa_1^3} \right\} \,\frac{1}{t^{\,3/2}};\,\,x \leq L.
\label{e11s}
\end{eqnarray}
Inserting Eq. (\ref{e11s}) into Eq. (\ref{e8}) allows to write the nonescape probability as
\begin{equation}
P(t) \approx P^e(t)+P^{e,ne}(t)+P^{ne}(t),
\label{pa}
\end{equation}
where $P^e(t)$ stands for the purely exponential decay contribution,
\begin{equation}
P^e(t)=|C_1|^2I_1 e^{-\Gamma_1 t/\hbar},
\label{pe}
\end{equation}
$P^{e,ne}(t)$ refers to the interference term that involves exponential and nonexponential contributions,
\begin{eqnarray}
P^{e,ne}(t) &=& -{\rm Re}\left\{\left[\frac{|C_1|^2I_1}{{\kappa^*_1}^3}-\frac{C_1^2Y_1}{\kappa_1^3} \right] b^*  e^{-i\mathcal{E}_1 t /\hbar}  \right\} \nonumber \\ [.3cm]
&& \times \, e^{-\Gamma_1 t/2\hbar } \frac{1}{t^{3/2}}
\label{pnee}
\end{eqnarray}
and $P^{ne}(t)$ stands for the long-time post-exponential contribution,
\begin{equation}
P^{ne}(t)=\frac{|b|^2}{2}{\rm Re}\left\{\left[\frac{|C_1|^2I_1}{|\kappa_1^3|^2}-\frac{C_1^2Y_1}{(\kappa_1^3)^2} \right] \right\}\frac{1}{t^{3}}.
\label{pne}
\end{equation}
In the above expressions $I_1$ y $Y_1$ are defined, respectively, as
\begin{equation}
I_1= \int_{0}^{L} |u_1(x)|^2 dx.
\label{in}
\end{equation}
and
\begin{equation}
Y_1= \int_{0}^{L} u^2_1(x) dx.
\label{yn}
\end{equation}

We end this Section by referring to an exact single-level resonance expression for the survival probability \cite{gcrr01}, that allows to derive an approximate expression for the time $t_0$, in lifetime units, for the transition from exponential to post-exponential decay,
\begin{equation}
t_0 \approx 5.41\,\ln (R)+12.25,
\label{t0}
\end{equation}
where we recall that $R=\mathcal{E}_1/\Gamma_1$. Equation (\ref{t0}) yields a good estimate of $t_0$ for values of $R \gtrsim 1$ which is more accurate than those given in Refs. \cite{winter61,peres80}.

\section{Models}
\begin{figure}[!tbp]
\rotatebox{0}{\includegraphics[width=3.0in]{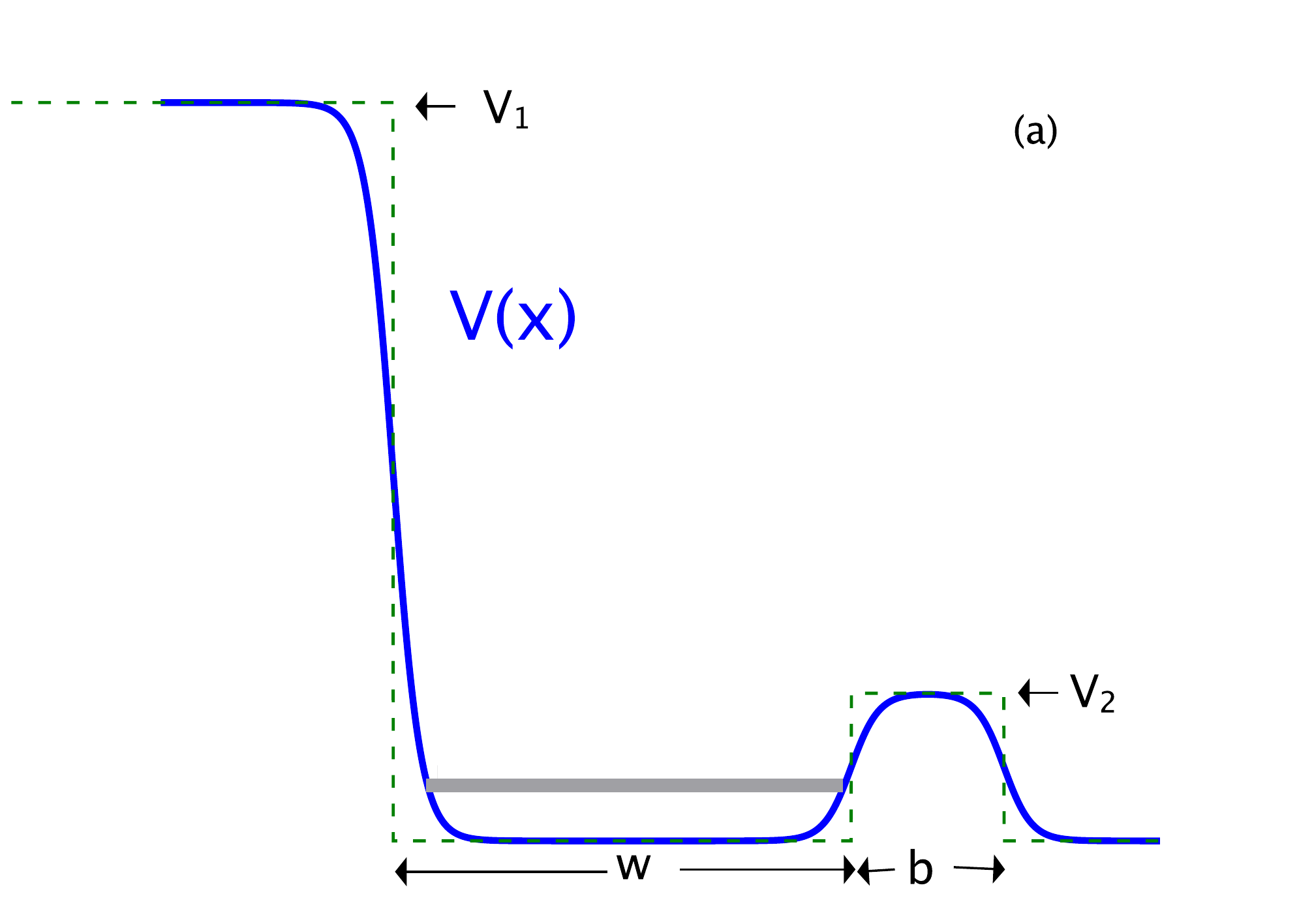}}
\rotatebox{0}{\includegraphics[width=3.0in]{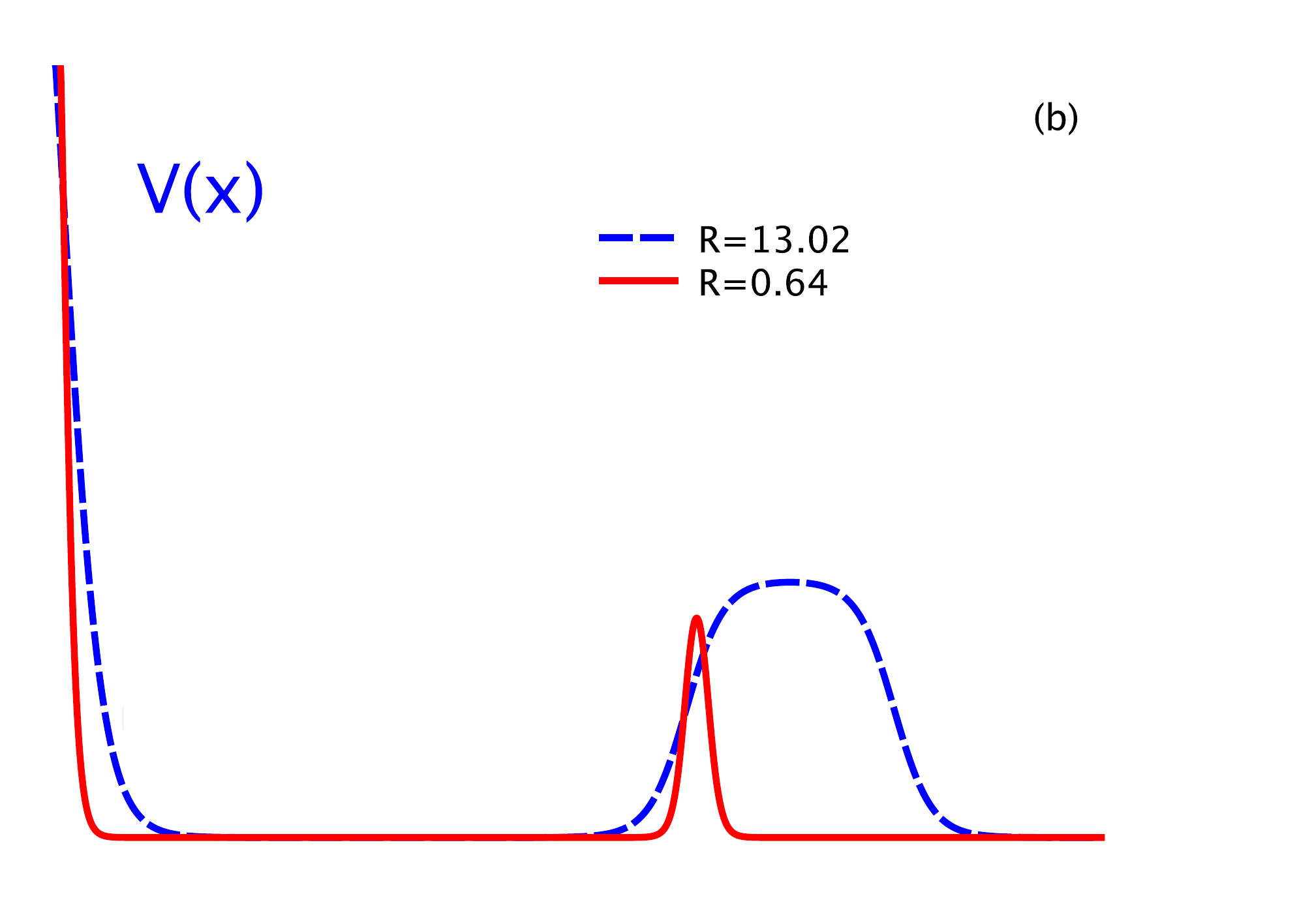}}
\caption{(Color online) (a) Trapping potential $V(x)$ (blue solid line) obtained by smoothing the squared barriers (green dashed line) using Eq.  (\ref{pot}).  (b) Two potential profiles appropriate for the analysis of the transition from exponential to post exponential decay. See parameters in the Table \ref{tabla1}.}
\label{figure1}
\end{figure}
\begin{figure}[!tbp]
\rotatebox{0}{\includegraphics[width=3.3in]{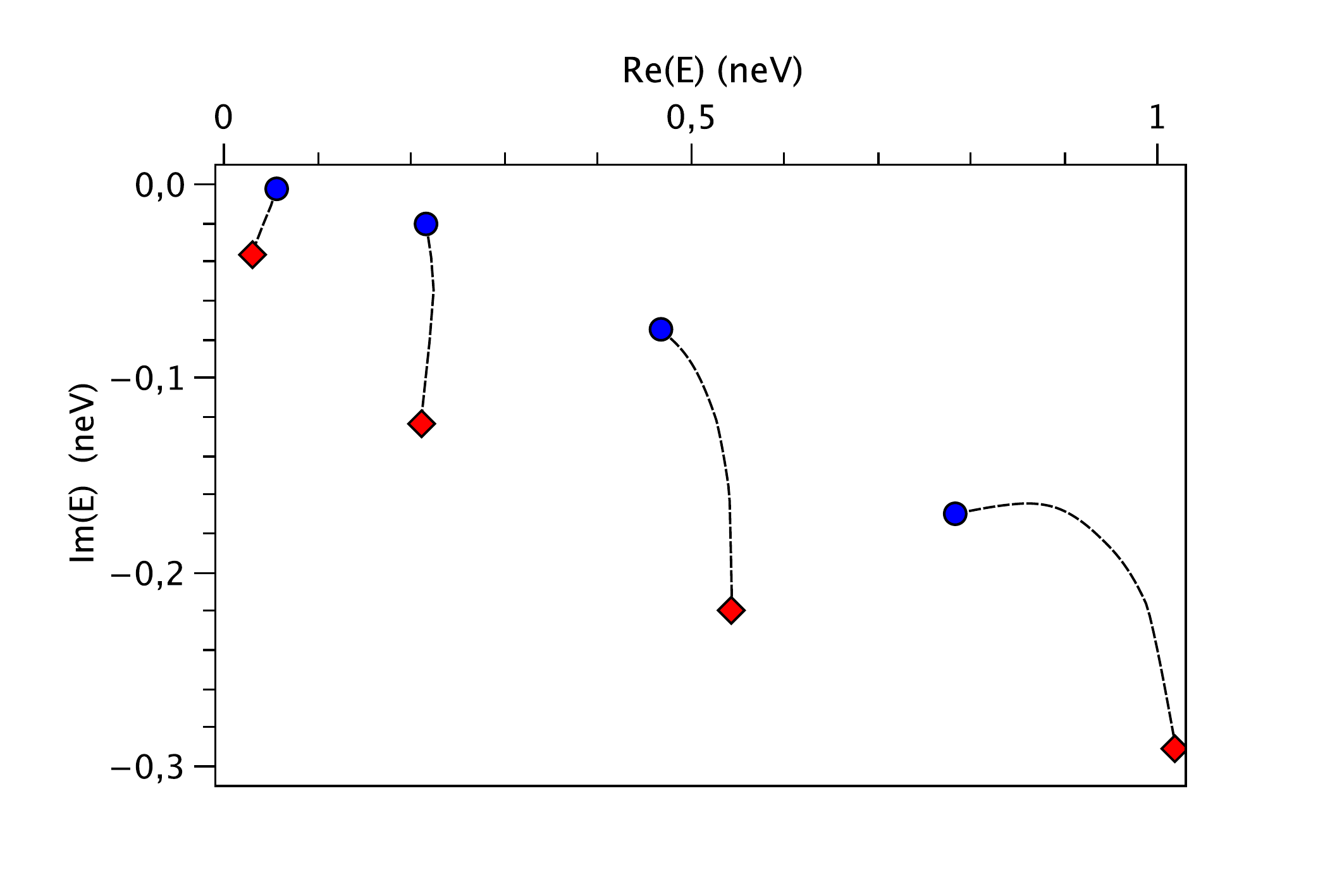}}
\caption{(Color online) Example of the trajectories followed by a few poles located on the complex energy plane as the tunneling barrier is gradually reduced. We consider neV units, where $1 {\rm neV}= 241.8$ ${\rm kHz}$. The poles in blue solid circles correspond to the potential profile depicted with a blue dashed line in Fig. \ref{figure1} (b) ($R=13.02$), whereas the poles in red diamonds refer to the potential with the thin tunneling barrier shown by the red solid line in  Fig. \ref{figure1} (b) ($R=0.64$).}
\label{figure2}
\end{figure}

As pointed out in the Introduction, a relevant aspect of the recent developments on the preparation of tunable few-body quantum systems is the control of the quantum states in these systems \cite{jochim11,zurn12,zurn13a}. For tunneling decay, the resulting potential corresponds to an impenetrable barrier on the left, a barrier of finite height on the right and a well in between that, in particular, it may be used to study tunneling decay of a single  atom located in the lowest energy level of the system \cite{jochim11}.

Here we consider two potential profiles  to study the  decaying regimes corresponding to different values of $R$ for single $^6{\rm Li}$ atom decay. The first potential profile is the bathtub potential, which has been considered in  Ref. \cite{pons12} to study multiparticle tunneling decay, and the second potential profile refers to the tunneling decay potential considered by the Heidelberg group on tunable few-fermion systems, which consists of the summation of the optical one-dimensional confinement potential plus a linear magnetic term \cite{jochim11}.

\subsection*{Bathtub potential}

Let us first refer to the bathtub potential. Figure \ref{figure1} (a) exhibits a profile for this potential (blue solid line), given by the formula
\begin{eqnarray}
V(x)=\frac{1}{2}V_1 \left[ \tanh{X_1} -  \tanh{X_2}   \right] \Theta \left(\frac{w}{2}-x\right) + \nonumber  \\
 \frac{1}{2}V_2 \left[ \tanh{X_1} -  \tanh{X_2}   \right] \Theta \left( x-\frac{w}{2} \right),\nonumber \\ [.1cm]
\label{pot}
\end{eqnarray}
where $\Theta(u)$ is the Heaviside step function, $X_1=  \left( |x-\frac{w}{2}|-\frac{w}{2} \right) / \sigma$, $X_2=  \left( x-L\right) / \sigma$, and $L=w+b$ being the total width of the potential depicted in Fig. \ref{figure1}(a) (green dashed line). The parameter $\sigma$ determines the smoothness of the potential. As required for single atom decay, the potential supports only one decaying state ($n=1$), with the higher states ($n \ge 2$) lying above the tunneling barrier height, \textit{i.e.}, $\mathcal{E}_n > V_2$ for $n=2,3,4,...$. The above may be accomplished by choosing appropriately the parameters of the potential shown in Fig. \ref{figure1}(a) and the value of $\sigma$. Here we fix for all calculations a large value for $V_1$, namely, $V_1/h=241.8\,{\rm kHz}$, so that the decay process occurs, as pointed out above, by tunneling through the second barrier $V_2$. Figure \ref{figure1}(b) displays two potential profiles which may be appropriate for the analysis of the distinct decaying regimes.

\begin{figure}[!tbp]
\rotatebox{0}{\includegraphics[width=3.1in]{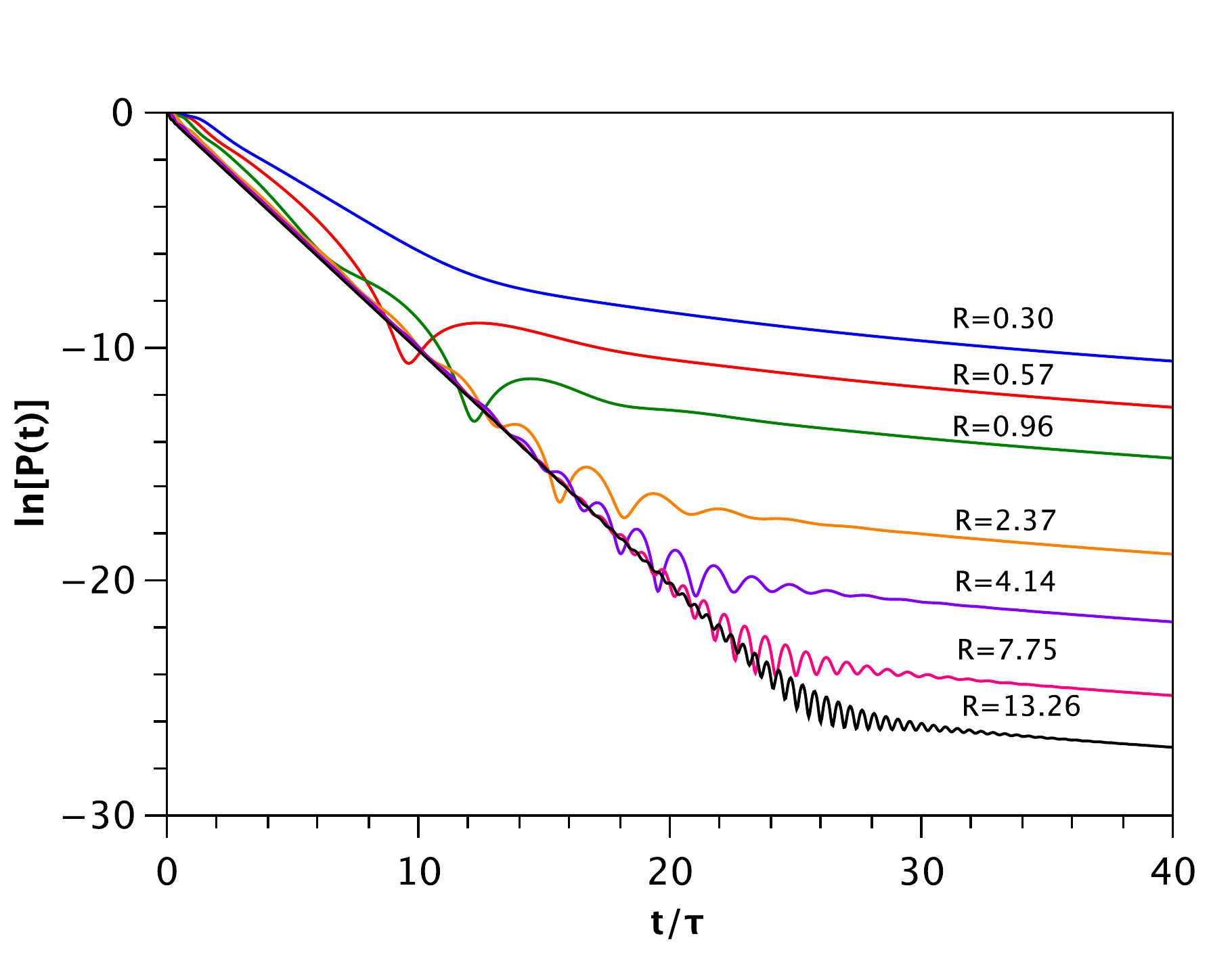}}
\caption{(Color online) Plot of the natural logarithm of the nonescape probability as a function of time (in units of the lifetime $\tau$) for different values of the ratio $R=\mathcal{E}_1/\Gamma_1$. The transition from the exponential to the post-exponential regime is clearly appreciated in each curve.}
\label{figure3}
\end{figure}

Using an appropriate combination of the system parameters may allow to select the transition time at which the decay changes from exponential to the post-exponential regime. According to Eq. (\ref{t0}), this transition time is tunable through variations of the ratio $R=\mathcal{E}_1/\Gamma_1$, which may be manipulated by realizing that the location of the poles on the complex wave number or energy planes depends on the system parameters.
The equation for the complex poles follows by imposing the outgoing boundary condition to Eq. (\ref{en1}). The poles may be easily calculated for potentials with rectangular shapes by well known procedures \cite{gcr97,cgc10a} that may be implemented to potentials of arbitrary shape by noticing that a given potential profile can be described as a sequence of rectangles of appropriate high and width.
As an example, Fig. \ref{figure2} shows that the main effect of diminishing the barrier width is to increase the values of the imaginary energies of the poles. One appreciates in Fig. \ref{figure2} the trajectories followed by some poles on the fourth quadrant of the energy plane (dashed line).

We model the initial state $\Psi(x,0)$ as the lowest energy state of a potential with infinite walls, namely,
\begin{equation}
\Psi(x,0)=\left(\frac{2}{w}\right)^{1/2} \sin{\left[\frac{\pi}{w}x\right]}.
\label{eini}
\end{equation}
We choose the initial state (\ref{eini}), in addition to its mathematical simplicity, because it has the essential physical ingredient that initially there must a large probability to find the particle within the interaction region.  The above initial state guarantees that ${\rm Re}\{C_1^2\}$ is the largest contribution to Eq. (\ref{e5}) that is the condition that justifies the single pole approximation in the expansion of the decaying wave function, as discussed in the previous Section.

Using the initial state given by Eq. (\ref{eini}) and the set of poles $\{\kappa_n\}$ and decaying states $\{u_n(r)\}$ corresponding to a given set of potential parameters, allows to calculate the nonescape probability given by Eq. (\ref{pfull}). Figure  \ref{figure3} exhibits a plot of  the time evolution of the nonescape probability as a function of time for different values of $R$ in units of the corresponding lifetime. One may clearly appreciate the transitions from  exponential to nonexponential behavior for distinct values of $R$. The potential parameters that correspond to the above systems for distinct values of $R$ are given in the Table \ref{tabla1}, which
in addition, displays the corresponding values of the lifetimes $\tau$ and the expansion coefficients ${\rm Re}\,\{C_1^2\}$. Notice that the barrier width $b$ is a relevant parameter to diminish the values of $R$.

\begin{table}[!tbp]
\caption{\label{tabla1} Potential parameters for the distinct values of $R$ shown in Fig. \ref{figure3}.  $V_e/h$ $({\rm kHz})$ is the effective potential height that results from the right hand side of Eq. (\ref{pot}), where  $V_2/h$ $({\rm kHz})$ is the potential height, $w$ $(\mu m)$ is the well width, $\sigma$ the smoothness of the potential, $b$ $(\mu m)$ is the barrier width, $R=\mathcal{E}_1/\Gamma_1$,  $\tau$ $(ms)$ the lifetime, and ${\rm Re}\,\{C_1^2\}$ the expansion the coefficient for $n=1$. The mass of $^6{\rm Li}$ is taken as $10,964.898\,m_e$, with $m_e$ the electron mass. See text.}
\begin{ruledtabular}
\begin{tabular}{cccccccc}
 $V_{e}/h $ & $V_2/h$& $w$ & $\sigma$  &  $b$ & $R$ & $\tau$ &${\rm Re}\,\{C_1^2\}$\\
   \hline
  4.19 & 5.56 &2.7 & 0.10 &0.53 &  13.26 & 1.844 & 0.849 \\
       \hline
  3.88 & 5.56 & 2.7&0.09 & 0.42 & 7.75 & 1.166 & 0.868 \\
       \hline
  3.60 & 6.04& 2.7  &0.07 & 0.26 & 4.14 & 0.719 & 0.908 \\
       \hline
  3.37 & 6.04& 2.7 &0.05 & 0.17 & 2.37 &  0.482 & 0.950 \\
       \hline
  2.23 & 6.04& 3.0 &0.03 & 0.07 & 0.96 & 0.308 & 1.079 \\
       \hline
  1.32 & 6.04& 3.0 &0.03 &0.04 &  0.57 & 0.223 & 1.214 \\
       \hline
  0.66 & 6.04& 3.0 &0.03 & 0.02 & 0.30 & 0.164 & 1.498 \\
\end{tabular}
\end{ruledtabular}
\end{table}
\begin{figure}[!tbp]
\rotatebox{0}{\includegraphics[width=3.3in]{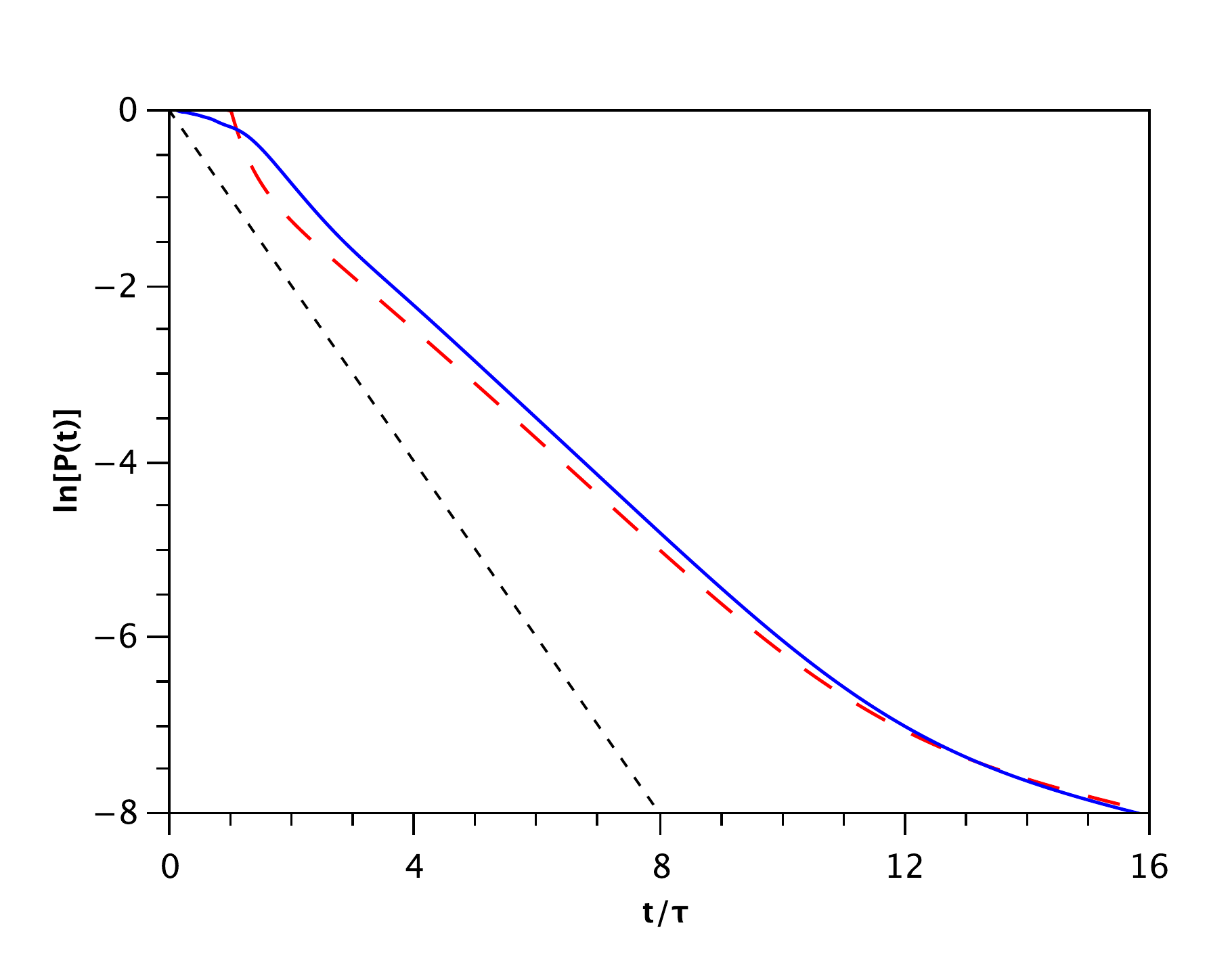}}
\caption{(Color online) Comparison of $\ln P(t)$ vs $t/\tau$ with $R=0.30$ calculated from the formal solution, Eq (\ref{pfull}) with $N=10$ resonant terms (solid line) and the values the sum $P^{e,ne}+P^{pne}$ (long dashed line) calculated from Eqs.  (\ref{pnee}) and  (\ref{pne}) respectively. The purely exponential contribution (dashed line) is included to help the eye.}
\label{figure4}
\end{figure}

It is worth emphasizing a number of features exhibited by Fig. \ref{figure3}. A first one is that the transition time is reduced as the values of $R$  decrease; a second one, is that the frequency of oscillations that arise from the interference between the purely exponential and the long-time inverse power contributions in (\ref{e11}) also diminishes as $R$ decreases; the third one is that for values of $R \lesssim 1$, the nonescape probability starts to exhibit a departure from purely exponential decay before the transition to the $t^{-3}$ long-time behavior occurs, and finally, as $R$ diminishes further, as exemplified by $R=0.30$, the decay becomes nonexponential at all times.

Previous studies involving distinct systems have shown that the single-pole approximation for the nonescape probability, given by Eq. (\ref{pa}), is an excellent approximation for $R > 1$ \cite{gcmv07}. As mentioned before, for values of $R$ much larger than unity the exponential decay law holds for many lifetimes and hence the long-time nonexponential contribution is very small.

On the other hand, it follows also from inspection of Fig. \ref{figure3}, that for values of $R \sim 1$, as $R=0.96$, or even with $R=0.57$, the nonescape probability  exhibits a clear departure from exponential decay just after a few lifetimes to then follow a
nonexponential  behavior as $t^{-3}$, according to Eq. (\ref{pne}).  This suggests  that in systems around these values of $R$ nonexponential decay could be amenable to experimental verification.

We have found that for values of $R < 1$, the single-pole approximation is still a good approximation. As pointed out above, the case $R =0.30$ is particularly interesting because it exhibits nonexponential decay in the full time interval. Figure \ref{figure4} provides a comparison of a calculation of the nonescape probability (solid line) using Eq. (\ref{pfull}), where 10 poles are sufficient to get convergence of the expansion, with the single-pole approximate calculation $P^{e,ne}+P^{ne}$ (long-dash), using, respectively, Eqs. (\ref{pnee}) and (\ref{pne}). One sees that the interference term $P^{e,ne}$, that goes as $\exp(-\Gamma_1 t/2\hbar)t^{-3/2}$, describes the decay for the first few lifetimes whereas the last term $P^{ne}$, that goes as $t^{-3}$, becomes the dominant contribution from approximately fourteen lifetimes onwards. The above suggests that systems with $R \lesssim 0.3$  could also be appropriate to verify experimentally nonexponential decay.

\begin{figure}[!tbp]
\rotatebox{0}{\includegraphics[width=3.3in]{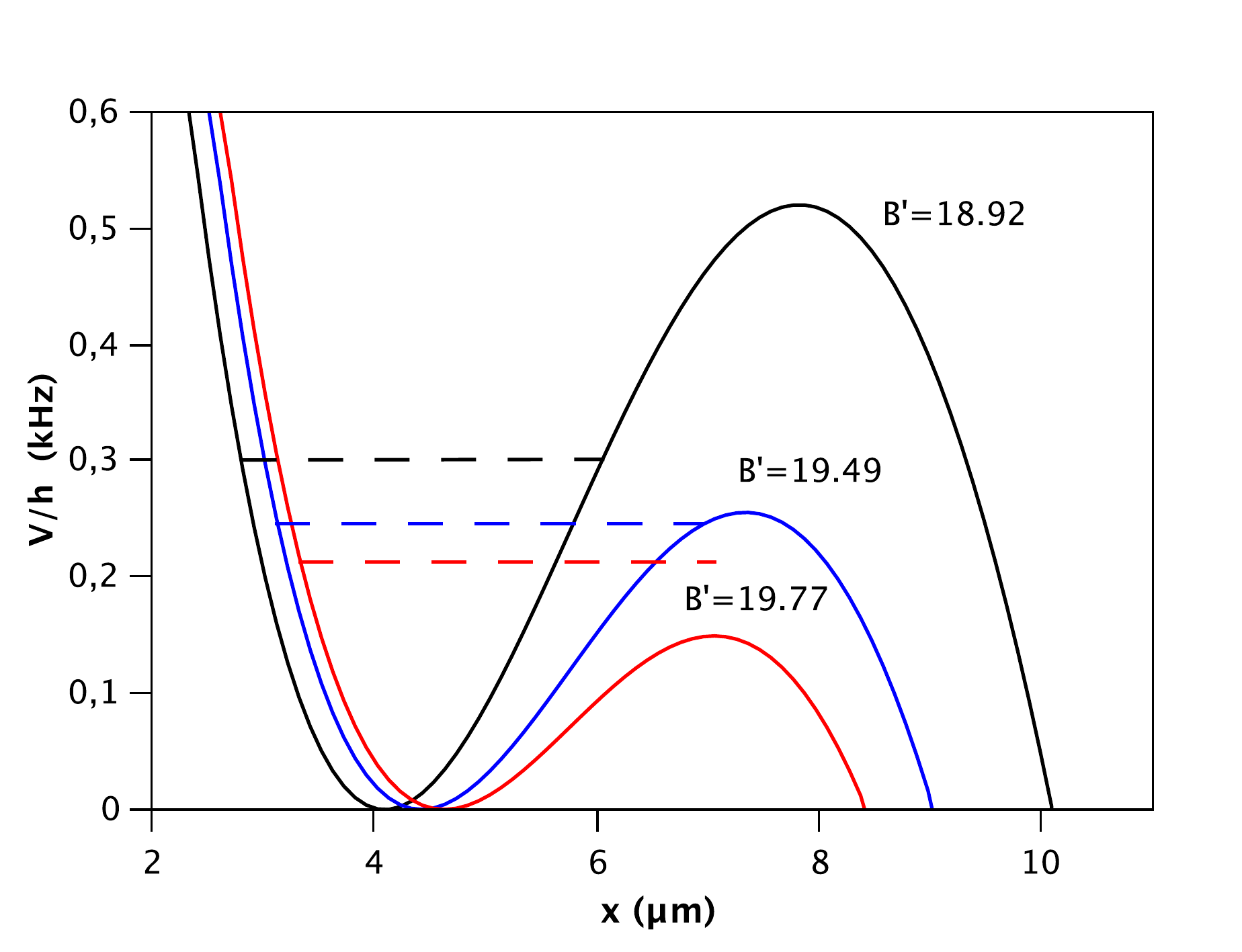}}
\caption{Color online) Potential profiles $V(x)$, given by  Eq. (\ref{pot2}), for three distinct values of the parameter $B^{\prime}$, as indicated in each curve. Each curve is shifted so that $V/h=0$ at the trap bottom. The dashed lines indicate the energies of the lowest decaying levels of each potential. The level with the largest energy corresponds to the potential with the highest barrier, and thus successively in descending order. See text.}
\label{figure5}
\end{figure}
\begin{figure}[!tbp]
\rotatebox{0}{\includegraphics[width=3.3in]{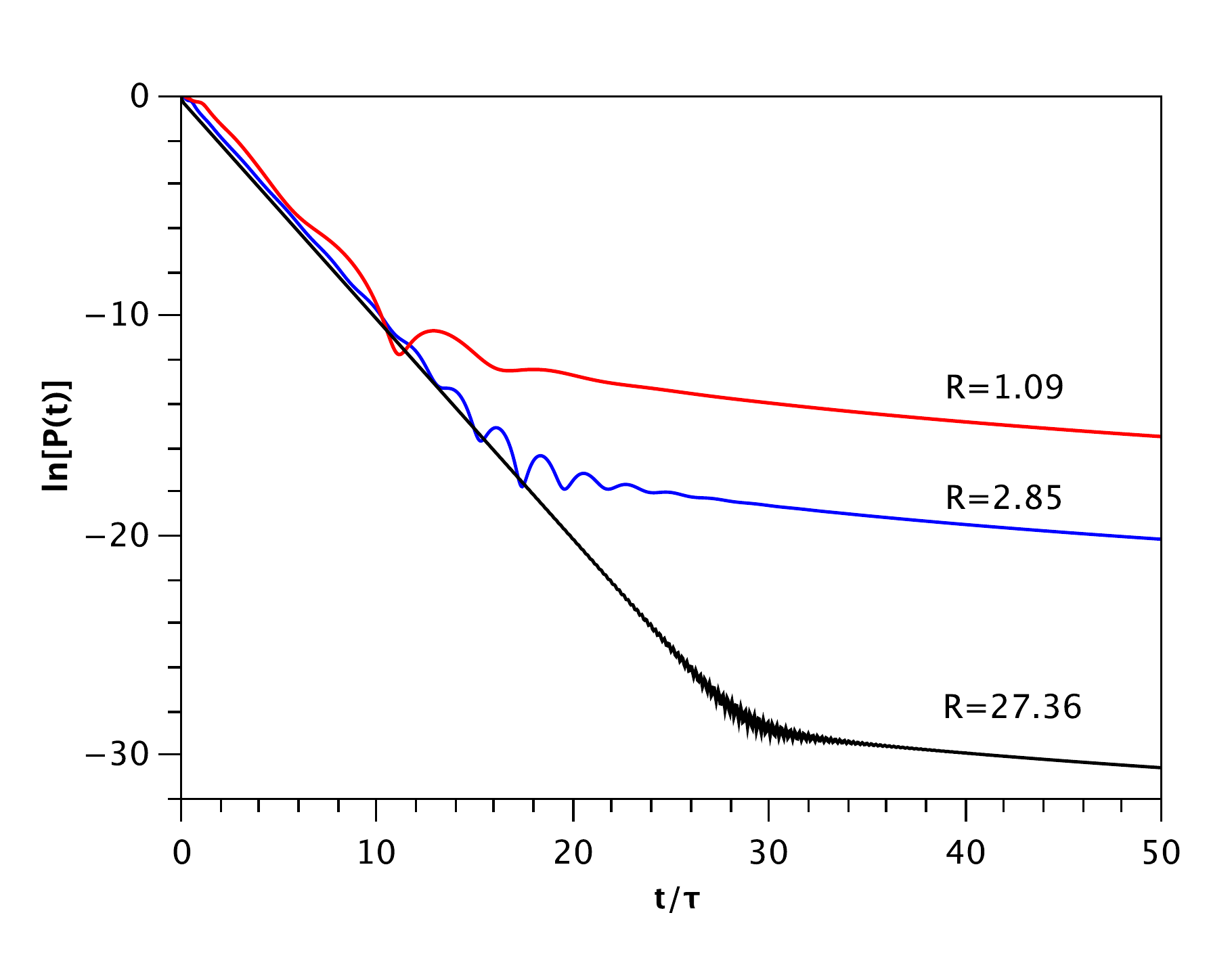}}
\caption{Color online) Plot of the natural logarithm of the nonescape probability $P(t)$ as a function of time in lifetime units $\tau$ for three different values of the $R=\mathcal{E}_1/\Gamma_1$, as indicated in each curve. See text.}
\label{figure6}
\end{figure}
\subsection*{Heidelberg potential}

Let us now refer to the second potential profile. This consists of a cigar-shaped cylindrically symmetric optical potential created by the sum of two terms, the first one a tightly focused laser beam which accounts by for an optical one-dimensional confinement of atoms and the second one, a linear magnetic potential term, that allows for tunneling decay. The potential is given by the formula \cite{jochim11,zurn13a,rontani13},
\begin{equation}
V(x)= p V_0 \left[1- \frac{1}{1+ (x/x_R)^2} \right] - {\mu}_m B^{\prime} x,
\label{pot2}
\end{equation}
where  $V_0=(3.326 \mu K) k_B$ is the initial depth at the center of the optical dipole trap, with $k_B$ the Boltzmann constant;  $p=0.6338$, the optical trap depth as a fraction of the initial depth; $x_R = \pi \omega_0^2/\lambda$ stands for the Rayleigh range, with $\lambda=1064 nm$,  the wavelength of the the trapping light; $\mu_m$ is the Bohr magneton; and $B^{\prime}=18.92\,G/cm$, is the magnetic field gradient \cite{jochim11,zurn13a,rontani13}. The above parameters determine a  value for $R$. The analysis by Rontani on single atom decay \cite{rontani13}, which is based on the experiments reported in Ref.  \cite{zurn13a}, gives a value of $R \approx 70$,  where
$\mathcal{E}_1= (316.3\,{\rm Hz})h$ and $\Gamma_1^{WKB}=(\gamma_{s0}/2\pi)=(4.516\,{\rm Hz})h$, with $h$ the Planck  constant and $\gamma_{s0}=(1/35.24) ms$ \cite{rontani13}. The above value of $R \approx 70$ yields using (\ref{t0}), an onset for nonexponential decay around $t_0  \approx 35$ lifetimes, which lies well beyond the range of $6$ lifetimes that were considered in these experiments. That analysis involves a trap parametrization involving a WKB analysis. In recent theoretical work, however, it is argued that the trap calibration via a WKB analysis leads to an inaccurate trap parametrization \cite{gharashi15}. Our own analysis, involving the complex pole for the above potential parameters,  gives $R=27.36$ (see below) which  gives an onset for nonexponential decay around $t_0 \approx 30$ lifetimes, that still lies beyond experimental verification.

It turns out that varying slightly the value of the magnetic field gradient $B^{\prime}$ modifies the potential profile. This is exemplified in Fig. \ref{figure5} which exhibits three potential profiles corresponding  respectively, to  $B^{\prime}=18.92$ $G/cm$, the case considered above,  $B^{\prime}=19.49$ $G/cm$ and $B^{\prime}=19.77$ $G/cm$, as indicated for each curve in that figure. Notice that we have shifted the origin of energy to the bottom of the potential for each case. The lowest energy decaying levels for each potential profile are indicated by  dashed lines in Fig. \ref{figure5}. The first level from above corresponds to the potential profile with $B^{\prime}=18.92$, and thus successively in descending order, as shown in Table \ref{tabla2}.

Figure \ref{figure6} displays the natural logarithm of the nonescape probability as a function of time in lifetime units for the above potential profiles corresponding, as indicated in the figure,  to values  of $R=1.09$, $2.85$ and $27.36$. In these calculations the box model initial state given by (\ref{eini}) is chosen to yield, as expected on physical grounds, values of ${\rm Re}\,\{C_1^2\}$ around unity. The largest value of $R$, namely, $R=27.36$,  corresponds to the potential profile with the smallest value of $B^{\prime}$ in Fig. \ref{figure5}, that is, $B^{\prime}=18.92$, and thus respectively for the other cases as indicated in Table \ref{tabla2}.  One sees, therefore,  that by varying slightly the values of the magnetic field gradient allows for the design of potential profiles with distinct values of  $R$ including values $R <1$. Presumably one might also vary some other parameters of the potential, as the optical trap depth $p$, in order to look for values of $R$ which might be adequate for the experimental verification of  nonexponential decay. Table \ref{tabla2} groups also some other relevant parameters for the calculations shown in Figs. \ref{figure5} and \ref{figure6}.

There is a feature that is worth pointing out here that results from our treatment concerning the potential given by Eq. (\ref{pot2}). It occurs for values of $R \lesssim 2.5$ and may be exemplified by the case with $R=1.09$ appearing in Fig. \ref{figure6} and refers to the fact  that the lowest energy decaying level has an energy that lies above the top of the corresponding potential barrier $(B^{\prime}=19.77)$, as exhibited in Fig. \ref{figure5}. The reason that the nonescape probability for this decaying level behaves in a similar fashion as for decaying levels that are located below the potential height, as for example for $R=0.96$ in Fig. \ref{figure3}, follows from the fact that both cases have a large value of the coefficient ${\rm Re}\,\{C_1^2\}$, as shown in Tables \ref{tabla1} and  \ref{tabla2}. This follows from the notion that the initial state overlaps strongly with the lowest decaying level. In these calculations the values of $w$  for the initial states are given in Table \ref{tabla2} and the  maxima of the corresponding probability densities are centered at the maxima of the corresponding probability decaying densities. Hence, it does not seem to matter if the lowest decaying level is located above or below the potential barrier height. The higher energy decaying levels, in addition that decay much faster, have very small values of the coefficients  ${\rm Re}\,\{C_n^2\}$, with $n=2,3,...$, as follows from Eq. (\ref{e5}), and hence do not play a relevant role in the decay process except at very small times. Clearly the above considerations lie beyond the WKB framework.

It is worth stressing that the behavior with time of the nonescape probability for the potentials considered here  is quite similar, as follows from a comparison between Figs. \ref{figure3} and \ref{figure6}. This suggests that what matters is the value of $R$ independently of the specific shape of the potential profile.

\begin{table}[!tbp]
\caption{\label{tabla2} Values of the magnetic field gradient $B^{\prime}$ $(G/cm)$, $w$ $(\mu m)$, $R=\mathcal{E}_1/\Gamma_1$, the energy of the decaying state $\mathcal{E}_1$ (kHz),  the lifetime $\tau$ $(ms)$, and the expansion coefficient for $n=1$, ${\rm Re}\,\{C_1^2\}$, corresponding to Figs. \ref{figure5} and \ref{figure6}. See text.}
\begin{ruledtabular}
\begin{tabular}{cccccc}
 $B^{\prime}$& $w$  & $R$ & $\mathcal{E}_1$  & $\tau$ &  ${\rm Re}\,\{C_1^2\}$\\
 \hline
  18.92 & 4.51  &27.36 & 0.300 & 14.48 & 0.831 \\
  \hline
  19.49 & 4.80   &2.85 & 0.246 & 1.848 & 0.978 \\
  \hline
  19.77 &  4.80  &1.09 & 0.214 & 0.817 & 1.206  \\
  \end{tabular}
\end{ruledtabular}
\end{table}
\section{Concluding Remarks \label{conclusions}}

The approach discussed in this work provides a consistent analytical framework to discuss exponential and nonexponential contributions to quantum decay. We have exemplified the above for two model calculation for the decay of ultracold atoms out of a  trap having a barrier, with realistic parameters. We have pointed out the relevance of the ratio of the energy of the decaying fragment to the decaying width, $R=\mathcal{E}_1/\Gamma_1$ to determine the decaying regime as a function of time, in particular values of $R <1$ or $R \sim 1$ to obtain a nonexponential behavior  of the decaying system within a few lifetimes. Here, it is important to stress the result that different combinations of potential parameters may lead to similar values of $R$. It is not crucial to know the precise analytical form of the potential. Essentially, the barrier height controls the number of decaying states within the well; the well width controls the energy value of the decaying state, and the barrier width, the value of decaying width.  From an experimental point of view in order to fix a value of $R$ requires to acquire control over these parameters.

We hope that the analysis presented here will stimulate  experimentalists interested in fundamental issues to look for the verification of the nonexponential contributions to quantum decay of ultracold atoms in these systems.

\begin{acknowledgments}
G.G-C. acknowledges the partial financial support of DGAPA-UNAM under grant IN111814.
\end{acknowledgments}

\begin{thebibliography}{57}%
\makeatletter
\providecommand \@ifxundefined [1]{%
 \@ifx{#1\undefined}
}%
\providecommand \@ifnum [1]{%
 \ifnum #1\expandafter \@firstoftwo
 \else \expandafter \@secondoftwo
 \fi
}%
\providecommand \@ifx [1]{%
 \ifx #1\expandafter \@firstoftwo
 \else \expandafter \@secondoftwo
 \fi
}%
\providecommand \natexlab [1]{#1}%
\providecommand \enquote  [1]{``#1''}%
\providecommand \bibnamefont  [1]{#1}%
\providecommand \bibfnamefont [1]{#1}%
\providecommand \citenamefont [1]{#1}%
\providecommand \href@noop [0]{\@secondoftwo}%
\providecommand \href [0]{\begingroup \@sanitize@url \@href}%
\providecommand \@href[1]{\@@startlink{#1}\@@href}%
\providecommand \@@href[1]{\endgroup#1\@@endlink}%
\providecommand \@sanitize@url [0]{\catcode `\\12\catcode `\$12\catcode
  `\&12\catcode `\#12\catcode `\^12\catcode `\_12\catcode `\%12\relax}%
\providecommand \@@startlink[1]{}%
\providecommand \@@endlink[0]{}%
\providecommand \url  [0]{\begingroup\@sanitize@url \@url }%
\providecommand \@url [1]{\endgroup\@href {#1}{\urlprefix }}%
\providecommand \urlprefix  [0]{URL }%
\providecommand \Eprint [0]{\href }%
\providecommand \doibase [0]{http://dx.doi.org/}%
\providecommand \selectlanguage [0]{\@gobble}%
\providecommand \bibinfo  [0]{\@secondoftwo}%
\providecommand \bibfield  [0]{\@secondoftwo}%
\providecommand \translation [1]{[#1]}%
\providecommand \BibitemOpen [0]{}%
\providecommand \bibitemStop [0]{}%
\providecommand \bibitemNoStop [0]{.\EOS\space}%
\providecommand \EOS [0]{\spacefactor3000\relax}%
\providecommand \BibitemShut  [1]{\csname bibitem#1\endcsname}%
\let\auto@bib@innerbib\@empty
\bibitem [{\citenamefont {Serwane}\ \emph {et~al.}(2011)\citenamefont
  {Serwane}, \citenamefont {Z\"urn}, \citenamefont {Lompe}, \citenamefont
  {Ottenstein}, \citenamefont {Wenz},\ and\ \citenamefont {Jochim}}]{jochim11}%
  \BibitemOpen
  \bibfield  {author} {\bibinfo {author} {\bibfnamefont {F.}~\bibnamefont
  {Serwane}}, \bibinfo {author} {\bibfnamefont {G.}~\bibnamefont {Z\"urn}},
  \bibinfo {author} {\bibfnamefont {T.}~\bibnamefont {Lompe}}, \bibinfo
  {author} {\bibfnamefont {T.}~\bibnamefont {Ottenstein}}, \bibinfo {author}
  {\bibfnamefont {A.~N.}\ \bibnamefont {Wenz}}, \ and\ \bibinfo {author}
  {\bibfnamefont {S.}~\bibnamefont {Jochim}},\ }\href@noop {} {\bibfield
  {journal} {\bibinfo  {journal} {Science}\ }\textbf {\bibinfo {volume}
  {332}},\ \bibinfo {pages} {336} (\bibinfo {year} {2011})}\BibitemShut
  {NoStop}%
\bibitem [{\citenamefont {Z\"urn}\ \emph {et~al.}(2012)\citenamefont {Z\"urn},
  \citenamefont {Serwane}, \citenamefont {Lompe}, \citenamefont {Wenz},
  \citenamefont {Ries}, \citenamefont {Bohn},\ and\ \citenamefont
  {Jochim}}]{zurn12}%
  \BibitemOpen
  \bibfield  {author} {\bibinfo {author} {\bibfnamefont {G.}~\bibnamefont
  {Z\"urn}}, \bibinfo {author} {\bibfnamefont {F.}~\bibnamefont {Serwane}},
  \bibinfo {author} {\bibfnamefont {T.}~\bibnamefont {Lompe}}, \bibinfo
  {author} {\bibfnamefont {A.~N.}\ \bibnamefont {Wenz}}, \bibinfo {author}
  {\bibfnamefont {M.~G.}\ \bibnamefont {Ries}}, \bibinfo {author}
  {\bibfnamefont {J.~E.}\ \bibnamefont {Bohn}}, \ and\ \bibinfo {author}
  {\bibfnamefont {S.}~\bibnamefont {Jochim}},\ }\href@noop {} {\bibfield
  {journal} {\bibinfo  {journal} {Phys. Rev. Lett.}\ }\textbf {\bibinfo
  {volume} {108}},\ \bibinfo {pages} {075303} (\bibinfo {year}
  {2012})}\BibitemShut {NoStop}%
\bibitem [{\citenamefont {Z\"urn}\ \emph {et~al.}(2013)\citenamefont {Z\"urn},
  \citenamefont {Wenz}, \citenamefont {Murmann}, \citenamefont {Bergschneider},
  \citenamefont {Lompe},\ and\ \citenamefont {Jochim}}]{zurn13a}%
  \BibitemOpen
  \bibfield  {author} {\bibinfo {author} {\bibfnamefont {G.}~\bibnamefont
  {Z\"urn}}, \bibinfo {author} {\bibfnamefont {A.~N.}\ \bibnamefont {Wenz}},
  \bibinfo {author} {\bibfnamefont {S.}~\bibnamefont {Murmann}}, \bibinfo
  {author} {\bibfnamefont {A.}~\bibnamefont {Bergschneider}}, \bibinfo {author}
  {\bibfnamefont {T.}~\bibnamefont {Lompe}}, \ and\ \bibinfo {author}
  {\bibfnamefont {S.}~\bibnamefont {Jochim}},\ }\href@noop {} {\bibfield
  {journal} {\bibinfo  {journal} {Phys. Rev. Lett.}\ }\textbf {\bibinfo
  {volume} {111}},\ \bibinfo {pages} {175302} (\bibinfo {year}
  {2013})}\BibitemShut {NoStop}%
\bibitem [{\citenamefont {Bloch}\ \emph {et~al.}(2008)\citenamefont {Bloch},
  \citenamefont {Dalibard},\ and\ \citenamefont {Zwerger}}]{bloch08}%
  \BibitemOpen
  \bibfield  {author} {\bibinfo {author} {\bibfnamefont {I.}~\bibnamefont
  {Bloch}}, \bibinfo {author} {\bibfnamefont {J.}~\bibnamefont {Dalibard}}, \
  and\ \bibinfo {author} {\bibfnamefont {W.}~\bibnamefont {Zwerger}},\
  }\href@noop {} {\bibfield  {journal} {\bibinfo  {journal} {Rev. Mod. Phys.}\
  }\textbf {\bibinfo {volume} {80}},\ \bibinfo {pages} {885} (\bibinfo {year}
  {2008})}\BibitemShut {NoStop}%
\bibitem [{\citenamefont {Lode}\ \emph {et~al.}(2012)\citenamefont {Lode},
  \citenamefont {Streltsov}, \citenamefont {Sakmann}, \citenamefont {Alon},\
  and\ \citenamefont {Cederbaum}}]{lode12}%
  \BibitemOpen
  \bibfield  {author} {\bibinfo {author} {\bibfnamefont {A.~U.}\ \bibnamefont
  {Lode}}, \bibinfo {author} {\bibfnamefont {A.~I.}\ \bibnamefont {Streltsov}},
  \bibinfo {author} {\bibfnamefont {K.}~\bibnamefont {Sakmann}}, \bibinfo
  {author} {\bibfnamefont {O.~E.}\ \bibnamefont {Alon}}, \ and\ \bibinfo
  {author} {\bibfnamefont {L.~S.}\ \bibnamefont {Cederbaum}},\ }\href@noop {}
  {\bibfield  {journal} {\bibinfo  {journal} {Proceedings of the National
  Academy of Sciences}\ }\textbf {\bibinfo {volume} {109}},\ \bibinfo {pages}
  {13521} (\bibinfo {year} {2012})}\BibitemShut {NoStop}%
\bibitem [{\citenamefont {Guan}\ \emph {et~al.}(2013)\citenamefont {Guan},
  \citenamefont {Batchelor},\ and\ \citenamefont {Lee}}]{guan13}%
  \BibitemOpen
  \bibfield  {author} {\bibinfo {author} {\bibfnamefont {X.-W.}\ \bibnamefont
  {Guan}}, \bibinfo {author} {\bibfnamefont {M.~T.}\ \bibnamefont {Batchelor}},
  \ and\ \bibinfo {author} {\bibfnamefont {C.}~\bibnamefont {Lee}},\
  }\href@noop {} {\bibfield  {journal} {\bibinfo  {journal} {Rev. Mod. Phys.}\
  }\textbf {\bibinfo {volume} {85}},\ \bibinfo {pages} {1633} (\bibinfo {year}
  {2013})}\BibitemShut {NoStop}%
\bibitem [{\citenamefont {del Campo}\ \emph {et~al.}(2006)\citenamefont {del
  Campo}, \citenamefont {Delgado}, \citenamefont {Garc\'\i{}a-Calder\'on},
  \citenamefont {Muga},\ and\ \citenamefont {Raizen}}]{cdgcmr06}%
  \BibitemOpen
  \bibfield  {author} {\bibinfo {author} {\bibfnamefont {A.}~\bibnamefont {del
  Campo}}, \bibinfo {author} {\bibfnamefont {F.}~\bibnamefont {Delgado}},
  \bibinfo {author} {\bibfnamefont {G.}~\bibnamefont {Garc\'\i{}a-Calder\'on}},
  \bibinfo {author} {\bibfnamefont {J.~G.}\ \bibnamefont {Muga}}, \ and\
  \bibinfo {author} {\bibfnamefont {M.~G.}\ \bibnamefont {Raizen}},\ }\href
  {\doibase 10.1103/PhysRevA.74.013605} {\bibfield  {journal} {\bibinfo
  {journal} {Phys. Rev. A}\ }\textbf {\bibinfo {volume} {74}},\ \bibinfo
  {pages} {013605} (\bibinfo {year} {2006})}\BibitemShut {NoStop}%
\bibitem [{\citenamefont {del Campo}(2011)}]{delcampo11}%
  \BibitemOpen
  \bibfield  {author} {\bibinfo {author} {\bibfnamefont {A.}~\bibnamefont {del
  Campo}},\ }\href@noop {} {\bibfield  {journal} {\bibinfo  {journal} {Phys.
  Rev. A}\ }\textbf {\bibinfo {volume} {84}},\ \bibinfo {pages} {012113}
  (\bibinfo {year} {2011})}\BibitemShut {NoStop}%
\bibitem [{\citenamefont {Longhi}\ and\ \citenamefont
  {Della~Valle}(2012)}]{longhi12}%
  \BibitemOpen
  \bibfield  {author} {\bibinfo {author} {\bibfnamefont {S.}~\bibnamefont
  {Longhi}}\ and\ \bibinfo {author} {\bibfnamefont {G.}~\bibnamefont
  {Della~Valle}},\ }\href@noop {} {\bibfield  {journal} {\bibinfo  {journal}
  {Phys. Rev. A}\ }\textbf {\bibinfo {volume} {86}},\ \bibinfo {pages} {012112}
  (\bibinfo {year} {2012})}\BibitemShut {NoStop}%
\bibitem [{\citenamefont {Pons}\ \emph {et~al.}(2012)\citenamefont {Pons},
  \citenamefont {Sokolovski},\ and\ \citenamefont {del Campo}}]{pons12}%
  \BibitemOpen
  \bibfield  {author} {\bibinfo {author} {\bibfnamefont {M.}~\bibnamefont
  {Pons}}, \bibinfo {author} {\bibfnamefont {D.}~\bibnamefont {Sokolovski}}, \
  and\ \bibinfo {author} {\bibfnamefont {A.}~\bibnamefont {del Campo}},\
  }\href@noop {} {\bibfield  {journal} {\bibinfo  {journal} {Phys. Rev. A}\
  }\textbf {\bibinfo {volume} {85}},\ \bibinfo {pages} {022107} (\bibinfo
  {year} {2012})}\BibitemShut {NoStop}%
\bibitem [{\citenamefont {Garc\'ia-Calder\'on}\ and\ \citenamefont
  {Mendoza-Luna}(2011)}]{gcm11}%
  \BibitemOpen
  \bibfield  {author} {\bibinfo {author} {\bibfnamefont {G.}~\bibnamefont
  {Garc\'ia-Calder\'on}}\ and\ \bibinfo {author} {\bibfnamefont {L.~G.}\
  \bibnamefont {Mendoza-Luna}},\ }\href@noop {} {\bibfield  {journal} {\bibinfo
   {journal} {Phys. Rev. A}\ }\textbf {\bibinfo {volume} {84}},\ \bibinfo
  {pages} {032106} (\bibinfo {year} {2011})}\BibitemShut {NoStop}%
\bibitem [{\citenamefont {Kim}\ and\ \citenamefont {Brand}(2011)}]{kim11}%
  \BibitemOpen
  \bibfield  {author} {\bibinfo {author} {\bibfnamefont {S.}~\bibnamefont
  {Kim}}\ and\ \bibinfo {author} {\bibfnamefont {J.}~\bibnamefont {Brand}},\
  }\href@noop {} {\bibfield  {journal} {\bibinfo  {journal} {Journal of Physics
  B: Atomic, Molecular and Optical Physics}\ }\textbf {\bibinfo {volume}
  {44}},\ \bibinfo {pages} {195301} (\bibinfo {year} {2011})}\BibitemShut
  {NoStop}%
\bibitem [{\citenamefont {Rontani}(2012)}]{rontani12}%
  \BibitemOpen
  \bibfield  {author} {\bibinfo {author} {\bibfnamefont {M.}~\bibnamefont
  {Rontani}},\ }\href@noop {} {\bibfield  {journal} {\bibinfo  {journal}
  {Physical Review letters}\ }\textbf {\bibinfo {volume} {108}},\ \bibinfo
  {pages} {115302} (\bibinfo {year} {2012})}\BibitemShut {NoStop}%
\bibitem [{\citenamefont {Rontani}(2013)}]{rontani13}%
  \BibitemOpen
  \bibfield  {author} {\bibinfo {author} {\bibfnamefont {M.}~\bibnamefont
  {Rontani}},\ }\href@noop {} {\bibfield  {journal} {\bibinfo  {journal} {Phys.
  Rev. A}\ }\textbf {\bibinfo {volume} {88}},\ \bibinfo {pages} {043633}
  (\bibinfo {year} {2013})}\BibitemShut {NoStop}%
\bibitem [{\citenamefont {Lundmark}\ \emph {et~al.}(2015)\citenamefont
  {Lundmark}, \citenamefont {Forss\'en},\ and\ \citenamefont
  {Rotureau}}]{lundmark15}%
  \BibitemOpen
  \bibfield  {author} {\bibinfo {author} {\bibfnamefont {R.}~\bibnamefont
  {Lundmark}}, \bibinfo {author} {\bibfnamefont {C.}~\bibnamefont {Forss\'en}},
  \ and\ \bibinfo {author} {\bibfnamefont {J.}~\bibnamefont {Rotureau}},\
  }\href {\doibase 10.1103/PhysRevA.91.041601} {\bibfield  {journal} {\bibinfo
  {journal} {Phys. Rev. A}\ }\textbf {\bibinfo {volume} {91}},\ \bibinfo
  {pages} {041601} (\bibinfo {year} {2015})}\BibitemShut {NoStop}%
\bibitem [{\citenamefont {Gharashi}\ and\ \citenamefont
  {Blume}(2015)}]{gharashi15}%
  \BibitemOpen
  \bibfield  {author} {\bibinfo {author} {\bibfnamefont {S.~E.}\ \bibnamefont
  {Gharashi}}\ and\ \bibinfo {author} {\bibfnamefont {D.}~\bibnamefont
  {Blume}},\ }\href@noop {} {\bibfield  {journal} {\bibinfo  {journal} {Phys.
  Rev. A}\ }\textbf {\bibinfo {volume} {92}},\ \bibinfo {pages} {033629}
  (\bibinfo {year} {2015})}\BibitemShut {NoStop}%
\bibitem [{\citenamefont {Garc\'{\i}a-Calder\'on}(2010)}]{gc10}%
  \BibitemOpen
  \bibfield  {author} {\bibinfo {author} {\bibfnamefont {G.}~\bibnamefont
  {Garc\'{\i}a-Calder\'on}},\ }\href@noop {} {\bibfield  {journal} {\bibinfo
  {journal} {Adv. Quant. Chem.}\ }\textbf {\bibinfo {volume} {60}},\ \bibinfo
  {pages} {407 } (\bibinfo {year} {2010})}\BibitemShut {NoStop}%
\bibitem [{\citenamefont {Garc\'\i{}a-Calder\'on}(2011)}]{gc11}%
  \BibitemOpen
  \bibfield  {author} {\bibinfo {author} {\bibfnamefont {G.}~\bibnamefont
  {Garc\'\i{}a-Calder\'on}},\ }\href@noop {} {\bibfield  {journal} {\bibinfo
  {journal} {AIP Conference Proceedings}\ }\textbf {\bibinfo {volume} {1334}},\
  \bibinfo {pages} {84} (\bibinfo {year} {2011})}\BibitemShut {NoStop}%
\bibitem [{\citenamefont {Garc\'ia-Calder\'on}\ \emph
  {et~al.}(2012)\citenamefont {Garc\'ia-Calder\'on}, \citenamefont {M\'attar},\
  and\ \citenamefont {Villavicencio}}]{gcmv12}%
  \BibitemOpen
  \bibfield  {author} {\bibinfo {author} {\bibfnamefont {G.}~\bibnamefont
  {Garc\'ia-Calder\'on}}, \bibinfo {author} {\bibfnamefont {A.}~\bibnamefont
  {M\'attar}}, \ and\ \bibinfo {author} {\bibfnamefont {J.}~\bibnamefont
  {Villavicencio}},\ }\href@noop {} {\bibfield  {journal} {\bibinfo  {journal}
  {Physica Scripta}\ }\textbf {\bibinfo {volume} {T151}},\ \bibinfo {pages}
  {014076} (\bibinfo {year} {2012})}\BibitemShut {NoStop}%
\bibitem [{\citenamefont {Khalfin}(1968)}]{khalfin68}%
  \BibitemOpen
  \bibfield  {author} {\bibinfo {author} {\bibfnamefont {L.~A.}\ \bibnamefont
  {Khalfin}},\ }\href@noop {} {\bibfield  {journal} {\bibinfo  {journal} {JETP
  Lett.}\ }\textbf {\bibinfo {volume} {8}},\ \bibinfo {pages} {65} (\bibinfo
  {year} {1968})}\BibitemShut {NoStop}%
\bibitem [{\citenamefont {Cordero}\ and\ \citenamefont
  {Garc\'\i{}a-Calder\'on}(2012)}]{cgc12}%
  \BibitemOpen
  \bibfield  {author} {\bibinfo {author} {\bibfnamefont {S.}~\bibnamefont
  {Cordero}}\ and\ \bibinfo {author} {\bibfnamefont {G.}~\bibnamefont
  {Garc\'\i{}a-Calder\'on}},\ }\href@noop {} {\bibfield  {journal} {\bibinfo
  {journal} {Phys. Rev. A}\ }\textbf {\bibinfo {volume} {86}},\ \bibinfo
  {pages} {062116} (\bibinfo {year} {2012})}\BibitemShut {NoStop}%
\bibitem [{\citenamefont {Misra}\ and\ \citenamefont
  {Sudarshan}(1977)}]{sudarshan77b}%
  \BibitemOpen
  \bibfield  {author} {\bibinfo {author} {\bibfnamefont {B.}~\bibnamefont
  {Misra}}\ and\ \bibinfo {author} {\bibfnamefont {E.~C.~G.}\ \bibnamefont
  {Sudarshan}},\ }\href@noop {} {\bibfield  {journal} {\bibinfo  {journal} {J.
  Math. Phys.}\ }\textbf {\bibinfo {volume} {18}},\ \bibinfo {pages} {756}
  (\bibinfo {year} {1977})}\BibitemShut {NoStop}%
\bibitem [{\citenamefont {Koshino}\ and\ \citenamefont
  {Shimizu}(2005)}]{koshino05}%
  \BibitemOpen
  \bibfield  {author} {\bibinfo {author} {\bibfnamefont {K.}~\bibnamefont
  {Koshino}}\ and\ \bibinfo {author} {\bibfnamefont {A.}~\bibnamefont
  {Shimizu}},\ }\href@noop {} {\bibfield  {journal} {\bibinfo  {journal} {Phys.
  Rep.}\ }\textbf {\bibinfo {volume} {412}},\ \bibinfo {pages} {191} (\bibinfo
  {year} {2005})}\BibitemShut {NoStop}%
\bibitem [{\citenamefont {Khalfin}(1958)}]{khalfin58}%
  \BibitemOpen
  \bibfield  {author} {\bibinfo {author} {\bibfnamefont {L.~A.}\ \bibnamefont
  {Khalfin}},\ }\href@noop {} {\bibfield  {journal} {\bibinfo  {journal} {Sov.
  Phys.--JETP}\ }\textbf {\bibinfo {volume} {6}},\ \bibinfo {pages} {1053}
  (\bibinfo {year} {1958})}\BibitemShut {NoStop}%
\bibitem [{\citenamefont {Goldberger}\ and\ \citenamefont
  {Watson}(1964)}]{goldberger64}%
  \BibitemOpen
  \bibfield  {author} {\bibinfo {author} {\bibfnamefont {M.~L.}\ \bibnamefont
  {Goldberger}}\ and\ \bibinfo {author} {\bibfnamefont {K.~M.}\ \bibnamefont
  {Watson}},\ }\href@noop {} {\emph {\bibinfo {title} {Collision Theory}}}\
  (\bibinfo  {publisher} {John Wiley and Sons, New York},\ \bibinfo {year}
  {1964})\BibitemShut {NoStop}%
\bibitem [{\citenamefont {Baz}\ \emph {et~al.}(1969)\citenamefont {Baz},
  \citenamefont {Zel'dovich},\ and\ \citenamefont {Perelomov}}]{baz69}%
  \BibitemOpen
  \bibfield  {author} {\bibinfo {author} {\bibfnamefont {A.~I.}\ \bibnamefont
  {Baz}}, \bibinfo {author} {\bibfnamefont {Y.~B.}\ \bibnamefont {Zel'dovich}},
  \ and\ \bibinfo {author} {\bibfnamefont {A.~M.}\ \bibnamefont {Perelomov}},\
  }\href@noop {} {\emph {\bibinfo {title} {Scattering, Reactions and Decay in
  Non-relativistic Quantum Merchanics}}}\ (\bibinfo  {publisher} {Israel
  Program for Scientific Translations, Jerusalem},\ \bibinfo {year}
  {1969})\BibitemShut {NoStop}%
\bibitem [{\citenamefont {Winter}(1961)}]{winter61}%
  \BibitemOpen
  \bibfield  {author} {\bibinfo {author} {\bibfnamefont {R.~G.}\ \bibnamefont
  {Winter}},\ }\href@noop {} {\bibfield  {journal} {\bibinfo  {journal} {Phys.
  Rev.}\ }\textbf {\bibinfo {volume} {123}},\ \bibinfo {pages} {1503} (\bibinfo
  {year} {1961})}\BibitemShut {NoStop}%
\bibitem [{\citenamefont {Fonda}\ \emph {et~al.}(1978)\citenamefont {Fonda},
  \citenamefont {Ghirardi},\ and\ \citenamefont {Rimini}}]{ghirardi78}%
  \BibitemOpen
  \bibfield  {author} {\bibinfo {author} {\bibfnamefont {L.}~\bibnamefont
  {Fonda}}, \bibinfo {author} {\bibfnamefont {G.~C.}\ \bibnamefont {Ghirardi}},
  \ and\ \bibinfo {author} {\bibfnamefont {A.}~\bibnamefont {Rimini}},\
  }\href@noop {} {\bibfield  {journal} {\bibinfo  {journal} {Rep. Prog. Phys.}\
  }\textbf {\bibinfo {volume} {41}},\ \bibinfo {pages} {587} (\bibinfo {year}
  {1978})}\BibitemShut {NoStop}%
\bibitem [{\citenamefont {Peres}(1980)}]{peres80}%
  \BibitemOpen
  \bibfield  {author} {\bibinfo {author} {\bibfnamefont {A.}~\bibnamefont
  {Peres}},\ }\href@noop {} {\bibfield  {journal} {\bibinfo  {journal} {Ann. of
  Phys.}\ }\textbf {\bibinfo {volume} {129}},\ \bibinfo {pages} {33} (\bibinfo
  {year} {1980})}\BibitemShut {NoStop}%
\bibitem [{\citenamefont {Norman}\ \emph {et~al.}(1988)\citenamefont {Norman},
  \citenamefont {Gazes}, \citenamefont {Crane},\ and\ \citenamefont
  {Bennett}}]{norman88}%
  \BibitemOpen
  \bibfield  {author} {\bibinfo {author} {\bibfnamefont {E.~B.}\ \bibnamefont
  {Norman}}, \bibinfo {author} {\bibfnamefont {S.~B.}\ \bibnamefont {Gazes}},
  \bibinfo {author} {\bibfnamefont {S.~G.}\ \bibnamefont {Crane}}, \ and\
  \bibinfo {author} {\bibfnamefont {D.~A.}\ \bibnamefont {Bennett}},\
  }\href@noop {} {\bibfield  {journal} {\bibinfo  {journal} {Phys. Rev. Lett.}\
  }\textbf {\bibinfo {volume} {60}},\ \bibinfo {pages} {2246} (\bibinfo {year}
  {1988})}\BibitemShut {NoStop}%
\bibitem [{\citenamefont {Nghiep}\ \emph {et~al.}(1998)\citenamefont {Nghiep},
  \citenamefont {Hanh},\ and\ \citenamefont {Son}}]{son98}%
  \BibitemOpen
  \bibfield  {author} {\bibinfo {author} {\bibfnamefont {T.~D.}\ \bibnamefont
  {Nghiep}}, \bibinfo {author} {\bibfnamefont {V.}~\bibnamefont {Hanh}}, \ and\
  \bibinfo {author} {\bibfnamefont {N.~N.}\ \bibnamefont {Son}},\ }\href@noop
  {} {\bibfield  {journal} {\bibinfo  {journal} {Nuclear Physics B (Proc.
  Suppl.)}\ }\textbf {\bibinfo {volume} {66}},\ \bibinfo {pages} {533}
  (\bibinfo {year} {1998})}\BibitemShut {NoStop}%
\bibitem [{\citenamefont {Nicolaides}\ and\ \citenamefont
  {Beck}(1977)}]{nicolaides77}%
  \BibitemOpen
  \bibfield  {author} {\bibinfo {author} {\bibfnamefont {C.~A.}\ \bibnamefont
  {Nicolaides}}\ and\ \bibinfo {author} {\bibfnamefont {D.~R.}\ \bibnamefont
  {Beck}},\ }\href@noop {} {\bibfield  {journal} {\bibinfo  {journal} {Phys.
  Rev. Lett.}\ }\textbf {\bibinfo {volume} {38}},\ \bibinfo {pages} {683}
  (\bibinfo {year} {1977})}\BibitemShut {NoStop}%
\bibitem [{\citenamefont {Parrott}\ and\ \citenamefont
  {Lawrence}(2002)}]{parrot02}%
  \BibitemOpen
  \bibfield  {author} {\bibinfo {author} {\bibfnamefont {R.~E.}\ \bibnamefont
  {Parrott}}\ and\ \bibinfo {author} {\bibfnamefont {J.}~\bibnamefont
  {Lawrence}},\ }\href@noop {} {\bibfield  {journal} {\bibinfo  {journal}
  {Europhys. Lett.}\ }\textbf {\bibinfo {volume} {57}},\ \bibinfo {pages} {632}
  (\bibinfo {year} {2002})}\BibitemShut {NoStop}%
\bibitem [{\citenamefont {Wilkinson}\ \emph {et~al.}(1997)\citenamefont
  {Wilkinson}, \citenamefont {Bharucha}, \citenamefont {Fischer}, \citenamefont
  {Madison}, \citenamefont {Morrow}, \citenamefont {Niu}, \citenamefont
  {Sundaram},\ and\ \citenamefont {Raizen}}]{raizen97}%
  \BibitemOpen
  \bibfield  {author} {\bibinfo {author} {\bibfnamefont {S.~R.}\ \bibnamefont
  {Wilkinson}}, \bibinfo {author} {\bibfnamefont {C.~F.}\ \bibnamefont
  {Bharucha}}, \bibinfo {author} {\bibfnamefont {M.~C.}\ \bibnamefont
  {Fischer}}, \bibinfo {author} {\bibfnamefont {K.~W.}\ \bibnamefont
  {Madison}}, \bibinfo {author} {\bibfnamefont {P.~R.}\ \bibnamefont {Morrow}},
  \bibinfo {author} {\bibfnamefont {Q.}~\bibnamefont {Niu}}, \bibinfo {author}
  {\bibfnamefont {B.}~\bibnamefont {Sundaram}}, \ and\ \bibinfo {author}
  {\bibfnamefont {M.~G.}\ \bibnamefont {Raizen}},\ }\href {\doibase
  10.1038/42418} {\bibfield  {journal} {\bibinfo  {journal} {Nature}\ }\textbf
  {\bibinfo {volume} {387}},\ \bibinfo {pages} {575} (\bibinfo {year}
  {1997})}\BibitemShut {NoStop}%
\bibitem [{\citenamefont {Fischer}\ \emph {et~al.}(2001)\citenamefont
  {Fischer}, \citenamefont {Guti\'errez-Medina},\ and\ \citenamefont
  {Raizen}}]{raizen01}%
  \BibitemOpen
  \bibfield  {author} {\bibinfo {author} {\bibfnamefont {M.~C.}\ \bibnamefont
  {Fischer}}, \bibinfo {author} {\bibfnamefont {B.}~\bibnamefont
  {Guti\'errez-Medina}}, \ and\ \bibinfo {author} {\bibfnamefont {M.~G.}\
  \bibnamefont {Raizen}},\ }\href@noop {} {\bibfield  {journal} {\bibinfo
  {journal} {Phys. Rev. Lett.}\ }\textbf {\bibinfo {volume} {87}},\ \bibinfo
  {pages} {040402} (\bibinfo {year} {2001})}\BibitemShut {NoStop}%
\bibitem [{\citenamefont {Rothe}\ \emph {et~al.}(2006)\citenamefont {Rothe},
  \citenamefont {Hintschich},\ and\ \citenamefont {Monkman}}]{monk06}%
  \BibitemOpen
  \bibfield  {author} {\bibinfo {author} {\bibfnamefont {C.}~\bibnamefont
  {Rothe}}, \bibinfo {author} {\bibfnamefont {S.~I.}\ \bibnamefont
  {Hintschich}}, \ and\ \bibinfo {author} {\bibfnamefont {A.~P.}\ \bibnamefont
  {Monkman}},\ }\href@noop {} {\bibfield  {journal} {\bibinfo  {journal} {Phys.
  Rev. Lett.}\ }\textbf {\bibinfo {volume} {96}},\ \bibinfo {pages} {163601}
  (\bibinfo {year} {2006})}\BibitemShut {NoStop}%
\bibitem [{\citenamefont {Jittoh}\ \emph {et~al.}(2005)\citenamefont {Jittoh},
  \citenamefont {Matsumoto}, \citenamefont {Sato}, \citenamefont {Sato},\ and\
  \citenamefont {Takeda}}]{jittoh05}%
  \BibitemOpen
  \bibfield  {author} {\bibinfo {author} {\bibfnamefont {T.}~\bibnamefont
  {Jittoh}}, \bibinfo {author} {\bibfnamefont {S.}~\bibnamefont {Matsumoto}},
  \bibinfo {author} {\bibfnamefont {J.}~\bibnamefont {Sato}}, \bibinfo {author}
  {\bibfnamefont {Y.}~\bibnamefont {Sato}}, \ and\ \bibinfo {author}
  {\bibfnamefont {K.}~\bibnamefont {Takeda}},\ }\href@noop {} {\bibfield
  {journal} {\bibinfo  {journal} {Phys. Rev. A}\ }\textbf {\bibinfo {volume}
  {71}},\ \bibinfo {pages} {012109} (\bibinfo {year} {2005})}\BibitemShut
  {NoStop}%
\bibitem [{\citenamefont {Garc\'ia-Calder\'on}\ and\ \citenamefont
  {Villavicencio}(2006)}]{gcv06}%
  \BibitemOpen
  \bibfield  {author} {\bibinfo {author} {\bibfnamefont {G.}~\bibnamefont
  {Garc\'ia-Calder\'on}}\ and\ \bibinfo {author} {\bibfnamefont
  {J.}~\bibnamefont {Villavicencio}},\ }\href@noop {} {\bibfield  {journal}
  {\bibinfo  {journal} {Phys. Rev. A}\ }\textbf {\bibinfo {volume} {73}},\
  \bibinfo {pages} {062115} (\bibinfo {year} {2006})}\BibitemShut {NoStop}%
\bibitem [{\citenamefont {Weisskopf}\ and\ \citenamefont
  {Wigner}(1930)}]{wigner30a}%
  \BibitemOpen
  \bibfield  {author} {\bibinfo {author} {\bibfnamefont {V.}~\bibnamefont
  {Weisskopf}}\ and\ \bibinfo {author} {\bibfnamefont {E.}~\bibnamefont
  {Wigner}},\ }\href@noop {} {\bibfield  {journal} {\bibinfo  {journal} {Z.
  Phys.}\ }\textbf {\bibinfo {volume} {63}},\ \bibinfo {pages} {54} (\bibinfo
  {year} {1930})}\BibitemShut {NoStop}%
\bibitem [{\citenamefont {Mostowski}\ and\ \citenamefont
  {W\'odkiewicz}(1973)}]{mostowski73}%
  \BibitemOpen
  \bibfield  {author} {\bibinfo {author} {\bibfnamefont {J.}~\bibnamefont
  {Mostowski}}\ and\ \bibinfo {author} {\bibfnamefont {K.}~\bibnamefont
  {W\'odkiewicz}},\ }\href@noop {} {\bibfield  {journal} {\bibinfo  {journal}
  {Bull. Acad. Polon. Sci.}\ }\textbf {\bibinfo {volume} {21}},\ \bibinfo
  {pages} {1027} (\bibinfo {year} {1973})}\BibitemShut {NoStop}%
\bibitem [{\citenamefont {Knight}(1977)}]{knight77}%
  \BibitemOpen
  \bibfield  {author} {\bibinfo {author} {\bibfnamefont {P.~L.}\ \bibnamefont
  {Knight}},\ }\href@noop {} {\bibfield  {journal} {\bibinfo  {journal} {Phys.
  Lett.}\ }\textbf {\bibinfo {volume} {61A}},\ \bibinfo {pages} {25} (\bibinfo
  {year} {1977})}\BibitemShut {NoStop}%
\bibitem [{\citenamefont {Gamow}(1928)}]{gamow28}%
  \BibitemOpen
  \bibfield  {author} {\bibinfo {author} {\bibfnamefont {G.}~\bibnamefont
  {Gamow}},\ }\href@noop {} {\bibfield  {journal} {\bibinfo  {journal} {Z.
  Phys.}\ }\textbf {\bibinfo {volume} {51}},\ \bibinfo {pages} {204} (\bibinfo
  {year} {1928})}\BibitemShut {NoStop}%
\bibitem [{\citenamefont {Gamow}\ and\ \citenamefont
  {Critchfield}(1949)}]{gamow49}%
  \BibitemOpen
  \bibfield  {author} {\bibinfo {author} {\bibfnamefont {G.}~\bibnamefont
  {Gamow}}\ and\ \bibinfo {author} {\bibfnamefont {C.~L.}\ \bibnamefont
  {Critchfield}},\ }\href@noop {} {\emph {\bibinfo {title} {Theory of Atomic
  Nucleus and Nucluer Energy--Sources}}}\ (\bibinfo  {publisher} {Oxford at the
  Clarendon Press},\ \bibinfo {year} {1949})\BibitemShut {NoStop}%
\bibitem [{\citenamefont {Berggren}(1968)}]{berggren68}%
  \BibitemOpen
  \bibfield  {author} {\bibinfo {author} {\bibfnamefont {T.}~\bibnamefont
  {Berggren}},\ }\href@noop {} {\bibfield  {journal} {\bibinfo  {journal}
  {Nucl. Phys. A}\ }\textbf {\bibinfo {volume} {109}},\ \bibinfo {pages} {265}
  (\bibinfo {year} {1968})}\BibitemShut {NoStop}%
\bibitem [{\citenamefont {Garc\'{i}a-Calder\'{o}n}\ and\ \citenamefont
  {Peierls}(1976)}]{gcp76}%
  \BibitemOpen
  \bibfield  {author} {\bibinfo {author} {\bibfnamefont {G.}~\bibnamefont
  {Garc\'{i}a-Calder\'{o}n}}\ and\ \bibinfo {author} {\bibfnamefont {R.~E.}\
  \bibnamefont {Peierls}},\ }\href@noop {} {\bibfield  {journal} {\bibinfo
  {journal} {Nucl. Phys. A}\ }\textbf {\bibinfo {volume} {265}},\ \bibinfo
  {pages} {443} (\bibinfo {year} {1976})}\BibitemShut {NoStop}%
\bibitem [{\citenamefont {Newton}(2002)}]{newtonchap12}%
  \BibitemOpen
  \bibfield  {author} {\bibinfo {author} {\bibfnamefont {R.~G.}\ \bibnamefont
  {Newton}},\ }\href@noop {} {\emph {\bibinfo {title} {Scattering Theory of
  Waves and Particles}}},\ \bibinfo {edition} {2nd}\ ed.\ (\bibinfo
  {publisher} {Dover Publications INC.},\ \bibinfo {year} {2002})\ \bibinfo
  {note} {chap. 12}\BibitemShut {NoStop}%
\bibitem [{\citenamefont {Abramowitz}\ and\ \citenamefont
  {Stegun}(1968)}]{abramowitzchap7}%
  \BibitemOpen
  \bibfield  {author} {\bibinfo {author} {\bibfnamefont {M.}~\bibnamefont
  {Abramowitz}}\ and\ \bibinfo {author} {\bibfnamefont {I.}~\bibnamefont
  {Stegun}},\ }\href@noop {} {\emph {\bibinfo {title} {Handbook of Mathematical
  Functions}}}\ (\bibinfo  {publisher} {Dover, N. Y.},\ \bibinfo {year}
  {1968})\ \bibinfo {note} {chap. 7}\BibitemShut {NoStop}%
\bibitem [{\citenamefont {Poppe}\ and\ \citenamefont {Wijers}(1990)}]{poppe90}%
  \BibitemOpen
  \bibfield  {author} {\bibinfo {author} {\bibfnamefont {G.~P.~M.}\
  \bibnamefont {Poppe}}\ and\ \bibinfo {author} {\bibfnamefont {C.~M.~J.}\
  \bibnamefont {Wijers}},\ }\href@noop {} {\bibfield  {journal} {\bibinfo
  {journal} {ACM Transactions on Mathematical Software}\ }\textbf {\bibinfo
  {volume} {16}},\ \bibinfo {pages} {38} (\bibinfo {year} {1990})}\BibitemShut
  {NoStop}%
\bibitem [{\citenamefont {Halliwell}\ and\ \citenamefont
  {Yearsley}(2013)}]{halliwell13}%
  \BibitemOpen
  \bibfield  {author} {\bibinfo {author} {\bibfnamefont {J.~J.}\ \bibnamefont
  {Halliwell}}\ and\ \bibinfo {author} {\bibfnamefont {J.~M.}\ \bibnamefont
  {Yearsley}},\ }\href@noop {} {\bibfield  {journal} {\bibinfo  {journal}
  {Phys. Rev. A}\ }\textbf {\bibinfo {volume} {87}},\ \bibinfo {pages} {022114}
  (\bibinfo {year} {2013})}\BibitemShut {NoStop}%
\bibitem [{\citenamefont {de~la Madrid}\ \emph {et~al.}(2005)\citenamefont
  {de~la Madrid}, \citenamefont {Garc\'{\i}a-Calder\'on},\ and\ \citenamefont
  {Muga}}]{mgcm05}%
  \BibitemOpen
  \bibfield  {author} {\bibinfo {author} {\bibfnamefont {R.}~\bibnamefont
  {de~la Madrid}}, \bibinfo {author} {\bibfnamefont {G.}~\bibnamefont
  {Garc\'{\i}a-Calder\'on}}, \ and\ \bibinfo {author} {\bibfnamefont {J.~G.}\
  \bibnamefont {Muga}},\ }\href@noop {} {\bibfield  {journal} {\bibinfo
  {journal} {Czech. J. Phys.}\ }\textbf {\bibinfo {volume} {55}},\ \bibinfo
  {pages} {1141} (\bibinfo {year} {2005})}\BibitemShut {NoStop}%
\bibitem [{\citenamefont {Garc\'{i}a-Calder\'{o}n}\ and\ \citenamefont
  {Rubio}(1997)}]{gcr97}%
  \BibitemOpen
  \bibfield  {author} {\bibinfo {author} {\bibfnamefont {G.}~\bibnamefont
  {Garc\'{i}a-Calder\'{o}n}}\ and\ \bibinfo {author} {\bibfnamefont
  {A.}~\bibnamefont {Rubio}},\ }\href@noop {} {\bibfield  {journal} {\bibinfo
  {journal} {Phys. Rev. A}\ }\textbf {\bibinfo {volume} {55}},\ \bibinfo
  {pages} {3361} (\bibinfo {year} {1997})}\BibitemShut {NoStop}%
\bibitem [{\citenamefont {del Campo}\ \emph {et~al.}(2009)\citenamefont {del
  Campo}, \citenamefont {Garc\'{\i}a-Calder\'on},\ and\ \citenamefont
  {Muga}}]{cgcm09}%
  \BibitemOpen
  \bibfield  {author} {\bibinfo {author} {\bibfnamefont {A.}~\bibnamefont {del
  Campo}}, \bibinfo {author} {\bibfnamefont {G.}~\bibnamefont
  {Garc\'{\i}a-Calder\'on}}, \ and\ \bibinfo {author} {\bibfnamefont
  {J.}~\bibnamefont {Muga}},\ }\href@noop {} {\bibfield  {journal} {\bibinfo
  {journal} {Physics Reports}\ }\textbf {\bibinfo {volume} {476}},\ \bibinfo
  {pages} {1} (\bibinfo {year} {2009})}\BibitemShut {NoStop}%
\bibitem [{\citenamefont {Berti}\ \emph {et~al.}(2009)\citenamefont {Berti},
  \citenamefont {Cardoso},\ and\ \citenamefont {Starinets}}]{starinets09}%
  \BibitemOpen
  \bibfield  {author} {\bibinfo {author} {\bibfnamefont {E.}~\bibnamefont
  {Berti}}, \bibinfo {author} {\bibfnamefont {V.}~\bibnamefont {Cardoso}}, \
  and\ \bibinfo {author} {\bibfnamefont {A.~O.}\ \bibnamefont {Starinets}},\
  }\href@noop {} {\bibfield  {journal} {\bibinfo  {journal} {Class. Quantum
  Grav.}\ }\textbf {\bibinfo {volume} {26}},\ \bibinfo {pages} {16301}
  (\bibinfo {year} {2009})}\BibitemShut {NoStop}%
\bibitem [{\citenamefont {Tolstikhin}(2008)}]{tolstikhin08}%
  \BibitemOpen
  \bibfield  {author} {\bibinfo {author} {\bibfnamefont {O.~I.}\ \bibnamefont
  {Tolstikhin}},\ }\href {\doibase 10.1103/PhysRevA.77.032711} {\bibfield
  {journal} {\bibinfo  {journal} {Phys. Rev. A}\ }\textbf {\bibinfo {volume}
  {77}},\ \bibinfo {pages} {032711} (\bibinfo {year} {2008})}\BibitemShut
  {NoStop}%
\bibitem [{\citenamefont {Garc\'{\i}a-Calder\'on}\ \emph
  {et~al.}(2007)\citenamefont {Garc\'{\i}a-Calder\'on}, \citenamefont
  {Maldonado},\ and\ \citenamefont {Villavicencio}}]{gcmv07}%
  \BibitemOpen
  \bibfield  {author} {\bibinfo {author} {\bibfnamefont {G.}~\bibnamefont
  {Garc\'{\i}a-Calder\'on}}, \bibinfo {author} {\bibfnamefont {I.}~\bibnamefont
  {Maldonado}}, \ and\ \bibinfo {author} {\bibfnamefont {J.}~\bibnamefont
  {Villavicencio}},\ }\href@noop {} {\bibfield  {journal} {\bibinfo  {journal}
  {Phys. Rev. A}\ }\textbf {\bibinfo {volume} {76}},\ \bibinfo {pages} {012103}
  (\bibinfo {year} {2007})}\BibitemShut {NoStop}%
\bibitem [{\citenamefont {Garc\'\i{}a-Calder\'on}\ \emph
  {et~al.}(2001)\citenamefont {Garc\'\i{}a-Calder\'on}, \citenamefont
  {Riquer},\ and\ \citenamefont {Romo}}]{gcrr01}%
  \BibitemOpen
  \bibfield  {author} {\bibinfo {author} {\bibfnamefont {G.}~\bibnamefont
  {Garc\'\i{}a-Calder\'on}}, \bibinfo {author} {\bibfnamefont {V.}~\bibnamefont
  {Riquer}}, \ and\ \bibinfo {author} {\bibfnamefont {R.}~\bibnamefont
  {Romo}},\ }\href@noop {} {\bibfield  {journal} {\bibinfo  {journal} {J. Phys
  A: Math. Gen.}\ }\textbf {\bibinfo {volume} {34}},\ \bibinfo {pages} {4155}
  (\bibinfo {year} {2001})}\BibitemShut {NoStop}%
\bibitem [{\citenamefont {Cordero}\ and\ \citenamefont
  {Garc\'{i}a-Calder\'{o}n}(2010)}]{cgc10a}%
  \BibitemOpen
  \bibfield  {author} {\bibinfo {author} {\bibfnamefont {S.}~\bibnamefont
  {Cordero}}\ and\ \bibinfo {author} {\bibfnamefont {G.}~\bibnamefont
  {Garc\'{i}a-Calder\'{o}n}},\ }\href@noop {} {\bibfield  {journal} {\bibinfo
  {journal} {J. Phys. A: Math. Theor.}\ }\textbf {\bibinfo {volume} {43}},\
  \bibinfo {pages} {185301} (\bibinfo {year} {2010})}\BibitemShut {NoStop}%
\end{thebibliography}
\end{document}